# A Decision Support System for daily scheduling and routing of home healthcare workers with a lunch break consideration

Ömer Öztürkoğlu[1], Gökberk Özsakallı[1], Syed Shah Sultan Mohiuddin Qadri[2]
[1] Business Administration, İzmir, Turkey, [2] Industrial Engineering, Çankaya University, Ankara Turkey

**ABSTRACT**

This study examines a home healthcare scheduling and routing problem (HHSRP) with a lunch break requirement. This problem especially consists of lunch break constraint for caregivers in addition to other typical features of the HHSRP in literature such as hard time windows constraints for both patients and caregivers, and patients' preferences. The objective is to minimize both travel distance in a route and unvisited patient (penalty) cost. For this NP-Hard problem, we developed an effective Adaptive Large Neighborhood Search algorithm to provide high-quality solutions in a short amount of time. We tested the proposed four variants of the algorithm with the selected problem instances from the literature. The algorithms provided nearly all optimal solutions for 30-patient problem instances in 12 seconds on averages. Additionally, they provided better solutions to 36 problem instances up to 36% improvement in some instance classes. Moreover, the improved solutions achieved to visit up to 10 more patients. The algorithms are also shown to be very robust due to its low coefficient variance of 0.3 on average. The algorithm also requires a very reasonable amount of time to generate solutions up to 54 seconds for solving 100-patient instances. A decision support system, namely Home Healthcare Decision Support System (HHCSS) was also designed to play a positive role in preventing the current COVID-19 global pandemic. The system employs the proposed ALNS algorithm to solve various instances of approximately generated COVID-19 patients' data from Turkey. The main aim of developing HHCSS is to support the administrative staff of home healthcare from the tedious task of scheduling and routing of caregivers and to increase service responsiveness.

## 1 INTRODUCTION

Nowadays, home healthcare services (HHS) have been becoming more popular services than before all around the world due to the growing aging population, increasing congestion and medical costs in hospitals. According to U.S. Labor's projections for 2014-2024, home health care services are expected to grow 60% with an additional 800 thousand new jobs (U.S. Labor of Department, 2015). It is becoming an important alternative to the conventional healthcare system. An alternative that provides better and customized services at low cost, not only for the elderly or disabled people but also for those who need help recovering from an illness or injury in a personalized environment after getting treatment at the hospital. Hence, this alternative system also offers savings to governments or institutions in healthcare expenses. Therefore, researchers have been increasing their attention to design efficient home healthcare systems and finding better ways to manage it by getting involved in decision-making processes at different phases of the system.

There are two main problems that need to be solved every day by decision-makers at home healthcare institutions at the operational phase. First, they should decide which caregivers serve to which patients considering caregivers' skills and specialties and patients' requirements. The second problem is to schedule and route of the assigned caregivers, which is known as home healthcare scheduling and routing problem (HHSRP) in literature. In recent years, these two problems are considered simultaneously with the help of Operations Research tools. Hence, in the HHSRP, with an aim to minimize travel and service costs, decision-makers try to find an efficient route and schedule of the caregivers simultaneously with their assignment to patients. The route of the caregivers starts from a health care center (HHC) and after providing the appropriate service to the patients again ends at the HHC. The mismanagement of this scheduling and routing problems may cause unserved and unsatisfied patients, high working times of caregivers, and high travel and service costs. According to Holm and Angelsen (2014), home health caregivers in Norway spend 22% of their time in vehicles, which can be accounted for as non-value added time. The home caregivers in the US travel nearly 8 billion miles each year (NAHC, 2015). Thus, the implementation of efficient solutions to the HHSRP may lead to huge cost savings to institutions and increase patients' satisfaction. However, in practice, we face with sub-optimal or inefficient solutions because of the personnel, who is in charge of solving the HHSRP, generally come from the medical background rather than an operations and information systems background. Therefore, this study aims to provide an efficient solution algorithm to one of the practical HHSRPs that also can easily be used in a decision support system (DSS) for managers to derive daily schedules easily.

Even though Fernandez et al. (1974) mentioned the problem of scheduling nurses in a rural area, the first detailed works of HHSRP dates back to late 1990s (Begur et al., 1997 and Cheng and Rich, 1998). Since that time, researchers have been developing models and solution algorithms for solving the variants of HHSRP with real-life or randomly generated data. Two comprehensive literature reviews related to HHSRP were published recently by Cissé et al. (2017) and Fikar and Hirsch (2017). In these review studies, the authors highlighted the characteristics of the variants of the HHSRPs such as time windows of patients and caregivers, caregivers' skill requirements, working times and break needs, temporal dependency such as synchronization of multiple services, precedent requirement of services, continuity of cares, workload balance of caregivers, single or multiple HHCs, different types of transportation modes, etc. Additionally, the previous studies related to the HHSRP were also categorized according to their objectives such as minimizing travel time or maximizing workload balance and, to proposed solution methods such as exact algorithms or metaheuristics. We let readers to review Cissé et al. (2017) and Fikar and Hirsch (2017) for the details of the previously published studies related to the HHSRP.

In compliance with the real-life practices, the majority of the previous studies also focused on a single-day planning horizon and single HHC in the HHSRP (Fikar and Hirsch, 2017). Similarly, the following three features or constraints are very commonly considered in the HHSRPs:

1. A patient's time window should be provided by a caregiver that defines the earliest and the latest start times of service.
2. A caregiver's qualifications should be matched to the patient's requirements. These qualifications could be defined as medical skills or expertise, language or social skills, or gender.
3. A working time regulation defines the maximum amount of time a caregiver is allowed to work in a shift, which is usually implemented by setting a time window.

In addition to the abovementioned features, there are a few numbers of rarely used constraints in the HHSRP such as

- Temporal dependency constraints consist of the synchronization of services, disjunctive services, and precedence constraints. In the synchronization of services, more than one caregiver has to appear for a simultaneous service. In disjunctive services, two services cannot be provided at the same time. And finally, precedence constraints, in which service could only be provided by a caregiver after another caregiver deliver a service. Dependency constraints consist of the synchronization of services, where more than one caregiver has to appear for a simultaneous service, or disjunctive services, where two services cannot be provided at the same time, and precedence constraints, in which a service could be provided by a caregiver after another caregiver deliver a service.
- Break constraints, which allow caregivers to have a rest or lunch within a specified time interval.

Among these constraints, break requirement is one of the most commonly seen practices in real-life because of both labor regulations and personnel needs.

Break requirement for caregivers was handled in different ways in previous studies. Some studies considered a predetermined time interval such as lunchtime and others provide a break after working a specified amount of time. For instance, Cheng and Rich (1998) and Villegas et al. (2018) allow workers to have a lunch break in the scheduled intervals before or after providing services in the vicinity of patients' locations rather than stopping at a location (node) on the way to the next patient. On the contrary, several studies such as Trautsamwieser and Hirsch (2011) and Wirnitzer et al. (2016) consider working time of caregivers to give them a break, which usually off 30 minutes after working 6 hours, at the break node, where it is visited before traveling to the patient. To the best of our knowledge, we reviewed all of the published journal articles and conference proceedings as well as unpublished online manuscripts that are related to the HHSRP with (lunch) break. These studies are briefly summarized according to the features of the HHSRP in Table 1. It can be seen in the table that the majority of the studies considered soft or hard time windows constraint for patients, medical expertise, fixed regular working time for caregivers, single HHC and single-period (daily) planning in their HHSRP. Moreover, only a few studies considered the synchronization of services among these studies (Eveborn et al.,2006; Bachouch et al., 2011; Xiao et al., 2018). Moreover, the majority of these studies solved their HHSRP for both real-life and randomly generated problem instances. Some studies used exact methods to find optimal route and schedule plan for caregivers (i.e. Bachouch et al., 2011; Liu et al., 2017), whereas most of them preferred to use metaheuristics for deriving good solutions due to the complexity of the HHSRP (i.e. Eveborn et al., 2006; Bertels and Fahle, 2006; Rest and Hirsch, 2016). Even though the majority of the studies dealt with real-life data for solving the HHSRP in order to demonstrate the usefulness of using Operations Research techniques in practice, they also used randomly generated instances to test their algorithm. However, most of the studies did not provide their problem instances, especially the real-life data due to confidentiality (see Table 1 for the accessibility of the problem instances).

In the light of the discussions, this study aims to deal with the HHSRP in which the aforementioned three common features and the lunch break constraint are considered because of their compliance with the practice. We then aim to develop an efficient algorithm for solving this problem. As a result, we choose the HHSRP proposed by Liu et al. (2017) because it complies with the practical and common features of the HHSRP. Additionally, Liu et al. (2017) used an exact method to derive optimal solutions for their publicly available randomly generated problem instances. Hence, we would compare the efficiency of our developed algorithm with the optimal or best-known solutions. Other than that, this paper also describes the Home Healthcare Decision

Support System (HHCSS) which is designed to support HHC center in a war against the COVID-19 pandemic. The developed Decision Support System employs the proposed ALNS algorithm to optimize the caregivers' itineraries of those instances generated using data of COVID-19 patients from the biggest cities of Turkey.

The remainder of this paper is structured as follows. Section 2 deals with the explanation of Liu et al. (2017)'s HHSRP and its mathematical model. The proposed Adaptive Large Neighborhood Search (ALNS) algorithm is presented in section 3. Section 4 encompasses the efficiency of the developed ALNS algorithm by comparing its solutions with Liu et al. (2017), the architecture HHCSS is explained in section 5. Experiments and empirical results on the data set of COVID-19 patients have been discussed in section 6 whereas the last section concludes the manuscript with our final remarks.

Table 1. The features of the HHSRP-LB in the literature.

| Reference | Patient's Time Window | Caregiver's Skills/Qualifications | Caregiver's Working Time | Break | # HHC | Temporal Dependency | Planning Horizon | Instance |
|---|---|---|---|---|---|---|---|---|
| Cheng and Rich (1998) | Hard | Medical Expertise | Fixed Regular Time + Overtime | Fixed Time Interval | Single HHC | NA | Single-period | Random (NA) & Real-life (NA) |
| Bertels and Fahle (2006) | Soft & Hard | Medical Expertise | Fixed Regular Time | Fixed Time Interval | Single HHC | NA | Single-period | Random (NA) |
| Eveborn et al. (2006) | Hard | Medical Expertise, Language, Gender | Fixed Regular Time + Overtime | Fixed Time Interval | Single HHC | Synchronized/ Shared | Single-period | Real-life (NA) |
| Bachouch et al. (2011) | Hard | Medical Expertise | Fixed Regular Time | Fixed Time Interval | Single HHC | Synchronized/ Shared | Multi-period | Random (NA) |
| Trautsamwieser and Hirsch (2011) | Soft & Hard | Medical Expertise | Fixed Regular Time + Overtime | After Working Amount of Time | Single HHC | NA | Single-period | Random (NA) & Real-life (NA) |
| Trautsamwieser et al. (2011) | Soft & Hard | Medical Expertise, Language | Fixed Regular Time + Overtime | After Working Amount of Time | Single HHC | NA | Single-period | Random (NA) & Real-life (NA) |
| Shao et al. (2012) | Soft & Hard | Medical Expertise | Fixed Regular Time | After Working Amount of Time | Multi HHC | NA | Multi-period | Random (NA) & Real-life (NA) |
| Bard et al. (2013) | NA | NA | Fixed Regular Time + Overtime | After Working Amount of Time | Multi HHC | NA | Multi-period | Real-life (NA) |
| Trautsamwieser and Hirsch (2014) | Hard | Medical Expertise | Fixed Regular Time | After Working Amount of Time | Single HHC | NA | Multi-period | Random (NA) & Real-life (NA) |
| Bard et al. (2014a) | Hard | Medical Expertise | Fixed Regular Time + Overtime | Fixed Time Interval | Multi HHC | NA | Multi-period | Random (NA) & Real-life (NA) |

Table 1. The features of the HHSRP-LB in the literature (Continued).

| Reference | Patient's Time Window | Caregiver's Skills/Qualifications | Caregiver's Working Time | Break | # HHC | Temporal Dependency | Planning Horizon | Instance |
|---|---|---|---|---|---|---|---|---|
| Bard et al. (2014b) | NA | NA | Fixed Regular Time + Overtime | After Working Amount of Time | Multi HHC | NA | Multi-period | Random based on Real-life (NA) |
| Fikar and Hirsch (2015) | Hard | Medical Expertise | Fixed Regular Time | After Working Amount of Time | Single HHC | NA | Single-period | Real-life (NA) |
| Rest and Hirsch (2016) | Hard | Medical Expertise, Language | Fixed Regular Time + Overtime | Fixed Time Interval | Single HHC | NA | Single-period | Real-life (NA) |
| Wirnitzer et al. (2016) | NA | Medical Expertise, Gender, Language, Preferences | Fixed Regular Time + Overtime | After Working Amount of Time | Single HHC | NA | Multi-period | Randomly from Real-life (NA) |
| Liu et al. (2017) | Hard | Medical Expertise | Fixed Regular Time | Fixed Time Interval | Single HHC | NA | Single-period | Random (A) & Real-life (NA) |
| Villegas et al. (2018) | NA | Medical Expertise | NA | Fixed Time Interval | Single HHC | NA | Single-period | Real-life (NA) |
| Xiao et al. (2018) | Hard | Medical Expertise, General Preference | Fixed Regular Time | Fixed Time Interval | Single HHC | Synchronized/ Shared | Single-period | Real-life (NA) |
| Liu et al. (2020) | Soft | Medical Expertise, General Preference | Fixed Regular Time | Fixed Time Interval | Multi HHC | NA | Multi-period | Random (NA) |

## 2   PROBLEM DEFINITION

As mentioned in Section 1, the home healthcare scheduling and routing problem with lunch break requirement, called HHSRP-LB hereafter, is considered in this study. Specifically, we studied the HHSRP-LB defined and formulated by Liu et al. (2017). This problem aims to find a set of routes of caregivers such that they leave from a HHC to serve patients, take a lunch break within their allowance and return to the HHC before the end of their shift. Additionally, it also aims to develop an efficient schedule of patients such that all of them are visited within their time preferences by the appropriate caregiver. Hence, the features of this problem could be described as in the following.

   i. Caregivers start and end their travel at the single HHC in their daily plan.
   ii. Caregivers have maximum working time in a day, therefore they must return to the HHC before the end of their time.
   iii. At the beginning of the day, all of the patients' requirements are known and caregivers are ready for a service at the HHC.
   iv. Each caregiver has his/her personal car to travel to patients.
   v. Patients have a hard time windows constraint such that caregivers should visit patients within their earliest and latest service start time interval, if possible.
   vi. Patients should be visited by an eligible or preferred caregiver.
   vii. The service time of a patient is deterministic and determined by the service requirement.
   viii. Travel times between patients are deterministic.
   ix. Caregivers must have a lunch break within the predetermined time interval and with a specified amount of duration.
   x. A penalty is incurred if a patient cannot be visited in the schedule due to patients' time windows and caregivers' working time constraints.

In the problem, the lunch break policy requires caregivers to decide whether they give a break before or after giving service to a patient considering the time windows constraints. Hence, this policy makes the problem even harder than a typical HHSRP. Additionally, lists of patients for each caregiver are described in order to maintain qualification requirements, continuity of care and to meet special preferences. Last, the objective of the model is described as minimizing the total traveling cost of caregivers and the total penalty cost of unvisited patients. Liu et al. (2017)'s formulation of the problem is given below and its notations were defined as in the following.

**Sets:**

$K$: Set of Available Workers, $K = \{1, 2, 3, \dots, m\}$.

$N$: Set of Clients, $N = \{1, 2, 3, \dots, n\}$.

$V$: Set of all nodes, $V = N \cup \{0, n+1\} = \{0, 1, 2, 3, \dots, n+1\}$.

$N_k$: Set indicating the clients whom worker $k \in K$ can serve

**Parameters:**

$d_i$: Service duration at client $i \in N$.

$[a_i, b_i]$: Time Window at client $i \in N$.

$a_i$: Earliest service time of client $i \in N$

$b_i$: Latest service time of client $i \in N$

$L$: Maximum working time of the worker.

$[0, L]$: Time Window of the worker.

$t_{ij}$: Travel time between node $i$ and node $j, i, j \in V$

$ct_{ij}$: Travel cost between node $i$ and node $j, i, j \in V$

$P$: Lunch break.

$B$: Lunch duration.

$[a_P, b_P]$: Time Window of lunch break.

$Q_{ik}$: 1, if worker $k \in K$ is allowed to serve client $i \in N$. 0, $otherwise$.

$cp_i$: Penalty cost if client $i \in N$ is not served.

**Decision Variables:**

$z_i$: 1, if $the$ client $i \in N$ is not visited. 0, otherwise

$x_{ijk}$: 1, if worker $k \in K$ $d$irectly visits client $j \in N$ $after$ $i \in N$. 0, otherwise.

$y_{ik}$: 1, if worker $k \in K$ takes a break at client $i \in N$ before service. 0, otherwise.

$y_{ik}'$: 1, if worker $k \in K$ takes a break at the client $i \in N$ after service. 0, otherwise.

$ts_{ik}$: Service start time of worker $k \in K$ at a client $i \in N$.

$ts_{Pk}$: Start time of the lunch break of a worker $k \in K$.

The mathematical model of Liu et al. (2017)'s HHSRP-LB

Objective: Min $\sum_{k \in K} \sum_{i \in V} \sum_{j \in V} ct_{ij} \, x_{ijk} + \sum_{i \in N} cp_i \, z_i$ (1)

Subject to:

$\sum_{k \in K} \sum_{j \in V} x_{ijk} + z_i = 1,$ $\forall \, i \in N$ (2)

$\sum_{j \in V} x_{0jk} = 1,$ $\forall \, k \in K$ (3)

$\sum_{j \in V} x_{j(n+1)k} = 1,$ $\forall \, k \in K$ (4)

$\sum_{i \in V} x_{ijk} - \sum_{i \in V} x_{jik} = 0,$ $\forall \, k \in K, j \in N$ (5)

$\sum_{i \in V} y_{ik} + \sum_{i \in V} y_{ik}' = 1,$ $\forall \, k \in K$ (6)

$y_{ik} + y_{ik}' \leq \sum_{j \in V} x_{ijk},$ $\forall \, k \in K, i \in V$ (7)

$ts_{ik} + (t_{ij} + d_i) x_{ijk} \leq ts_{jk} + (1 - x_{ijk}) b_i,$ $\forall \, k \in K, i, j \in V$ (8)

$ts_{Pk} + B \cdot y_{jk} \leq ts_{jk} + (1 - y_{jk}) b_P,$ $\forall \, k \in K, j \in V$ (9)

$ts_{ik} + (t_{ij} + d_i)(x_{ijk} + y_{jk} - 1) \leq ts_{Pk} + (2 - x_{ijk} - y_{jk}) b_i,$ $\forall \, k \in K, i, j \in V$ (10)

$$ts_{Pk} + (B + t_{ij})(x_{ijk} + y_{ik}' - 1) \leq ts_{jk} + (2 - x_{ijk} - y_{jk}')b_P, \quad \forall k \in K, i,j \in V \quad (11)$$

$$ts_{ik} + d_i y_{ik}' \leq ts_{Pk} + (1 - y_{ik}')b_i, \quad \forall k \in K, i \in V \quad (12)$$

$$a_i \sum_{j \in V} x_{ijk} \leq ts_{ik} \leq b_i \sum_{j \in V} x_{ijk}, \quad \forall k \in K, i \in N \quad (13)$$

$$0 \leq ts_{ik} \leq L, \quad k \in K, i \in \{0, n+1\} \quad (14)$$

$$a_P \leq ts_{Pk} \leq b_P, \quad \forall k \in K \quad (15)$$

$$x_{ijk} \leq \min\{Q_{ik}, Q_{ik}\}, \quad \forall k \in K, i,j \in V \quad (16)$$

$$y_{ik}, y_{ik}' \leq Q_{ik}, \quad \forall k \in K, i \in V \quad (17)$$

$$x_{ijk} \in \{0,1\}, \quad \forall k \in K, i,j \in V \quad (18)$$

$$y_{ik}, y_{ik}' \in \{0,1\}, \quad \forall k \in K, i \in V \quad (19)$$

$$z_i \in \{0,1\}, \quad \forall i \in N \quad (20)$$

In the given mathematical formulation of the HHSRP-LB, equation (2) ensures whether a patient is visited by a caregiver. Equations (3)-(5) are the network flow constraints that also forces to start from and travel back to the single HHC. Equations (6)-(7) are the lunch break constraints. Equation (8) is related to the feasibility of the route according to the service start times. Equations (9)-(12) also ensure the determination of caregivers' lunch break times. Equations (13)-(15) guarantee time windows constraints. Last, equations (18)-(20) define the variables' feasible values. For the details of the model, we let readers review Liu et al. (2017).

## 3 SOLUTION ALGORITHM

In the simplest form, the HHSRP consists of many features of a variant of a well-known Vehicle Routing Problem with Time Windows (VRPTW). Simply, both determine the sequence of patients/customers to be visited by caregivers/vehicles while minimizing the total traveling distance/cost. It is known that VRPTW has NP-Hard complexity (Lenstra and Kan, 1981). Therefore, we can conclude that HHSRP-LB has also the NP-Hard complexity.

Liu et al. (2017) developed a branch and price algorithm to optimally solve the problem. In their approach, the problem was divided into two parts. The set covering problem was considered in the first part as the master problem. In the second part, a sub-pricing problem that deals with shortest path problems with time windows and lunch break constraints were formulated. The authors developed labeling and Tabu Search algorithm to solve sub-pricing problems. Using randomly generated 168 problem instances, the authors showed that the developed branch and price algorithm is more effective than CPLEX commercial solver although it obtains optimal solutions for only 70% of the problem instances (48 unsolved instances). The authors also presented that the labeling algorithm is very time consuming that increases the run time of the algorithm. Thus, we propose an Adaptive Large Neighborhood Search (ALNS) algorithm in order to solve HHSRP-LB efficiently with optimal or close-optimal solutions in a short amount of time.

ALNS is one of the most commonly used metaheuristics algorithm for solving vehicle routing problems (VRP) in recent years. The algorithm was proposed and used by Ropke and Pisinger (2006a, 2006b). Simply, the algorithm searches for the optimal solution by ruining a solution and then repairing it iteratively. Hence, the algorithm consists of various removal and insertion heuristics for the purpose of ruin and repair, respectively. After an initial solution is obtained, the algorithm selects a pair of removal and insertion heuristics from a list of heuristics and applies them to the current solution to find a new solution at each iteration until the stopping criteria are met. Candidate solutions at each iteration are rejected or accepted according to a probabilistic simulated annealing criterion. The pseudocode of the proposed ALNS algorithm is presented in Table 2. The following sections explain each algorithmic step and the selected set of removal and insertion heuristics in detail.

Route of a caregiver is represented as a vector that contains depot, lunch break and patient nodes. For example, the route of caregiver $k$ is represented as $\pi_k = \{v_0, v_1, \ldots, v_b, v_i, \ldots, v_n, v_{n+1}\}$ where $v_0$ and $v_{n+1}$ are depots, $v_b$ is a lunch break node, and $v_i$ represent $n$ patient nodes. At the beginning of the algorithm, each route of caregivers contains depot nodes and a lunch break node. It is not allowed to remove these three nodes from the solution during the search process. Then, a solution $s$ is represented as the collection of routes of $m$ caregivers such as $s = \{\pi_1, \ldots, \pi_k, \ldots, \pi_m\}$.

Table 2. Pseudocode of the proposed ALNS algorithm A0.

**Inputs:** $\theta$, total number of iterations. $\bar{\theta}$, additional iterations. $\omega$ update solution iteration. $\tau_{Or}$, Or-opt local search iteration. $\tau_{Break}$, break local search iteration. $t$, current iteration.

generate an initial solution $s_{init}$
set $s_{best} \coloneqq s_{init}$ and $s_{curr} \coloneqq s_{init}$
**while** $t \leq \theta$ and $s_{best}$ has not improved last $\bar{\theta}$ iterations
    **if** $s_{best}$ has not improved last $\omega$ iterations **then**
        apply *Random Removal* and *Regret-3 Insertion* to $s_{curr}$ to generate $s_{new}$
    **else**
        select a *Removal Heuristic* randomly and apply to $s_{curr}$ to generate $s_{new}$
        select an *Insertion Heuristic* randomly and apply to $s_{new}$
    **end if**
    **if** $(t \, \% \, \tau_{Or}) = 0$ **then**
        apply *Or-opt heuristic* to $s_{curr}$
    **end if**
    **if** $(t \, \% \, \tau_{Break}) = 0$ **then**
        apply *Break local search heuristic* to $s_{curr}$
    **end if**
    **if** cost $f(s_{new})$ meets the *Acceptance Criteria* **then**
        $s_{curr} \coloneqq s_{new}$
        **if** $f(s_{new}) \leq f(s_{best})$ **then**
            $s_{best} \coloneqq s_{new}$
        **end if**
    **end if**
    $t = t + 1$
**end while**

**Output:** Best solution, $s_{best}$

## 3.1 Initial Solution

The initial solution is found by applying *regret-3* heuristic with noise which is explained in the insertion heuristic i.e. section 3.3 in detail. The idea of using *regret-3* heuristic is trying to escape from greedy approach that may increase the objective value by assigning latest nodes to worse positions in the routes (Perttunen, 1994; Van Breedam, 2001; Sousa et al., 2016). In addition, incorporating noise function to the heuristic can give more information about the robustness of the algorithm among replications. The reason is that each replication of an instance starts from a different initial solution with noise function. If the best solutions are similar among replications, it can be said that the algorithm is robust with respect to changing initial solutions.

At the beginning of the algorithm, all of the patients are placed into the request bank which consists of a set of patients that have not assigned to any route of caregiver yet. *Regret-3* with noise heuristic is applied to all of the available routes in parallel. If any patient cannot be assigned to any route, it remains in the request bank. Once a feasible solution is found, it is set to current and the best solution.

## 3.2 Removal Heuristics

At each iteration, a randomly selected removal heuristic removes a predetermined $q$ number of patients from the current solution $s_{curr}$ and places them to the request bank. In this study, we apply six (6) most commonly used removal heuristics for solving VRP using in ALNS algorithm: random, worst, Shaw, proximity, time and route removals (Ropke and Pisinger, 2006b; Pisinger and Ropke, 2007; Hemmelmayr et al., 2012; Grangier et al.. 2016; Koç et al., 2019).

**Random Removal:** Random removal heuristic is a simple heuristic that provides randomization to the search process. The heuristic selects $q$ number of patients from the current solution $s_{curr}$ at random, removes them from the solution and adds them to the request bank.

**Worst Removal:** Worst removal heuristic selects $q$ number of patients, which are the most costly patients in terms of distance, from the current. The heuristics then removes the selected patients from the current solution and adds them to the request bank.

**Shaw Removal:** Shaw removal heuristic was proposed by Shaw (1997, 1998). The main objective of the heuristic is to remove the most similar patients in terms of location and time windows. In doing this, the possibility of assigning the nearest patients to the same caregiver could be increased. Hence, the caregiver's route length may be decreased.

Shaw removal heuristic first removes a patient randomly from the current solution and adds it to the request bank. Next, a patient is randomly selected from the request bank. Similarities between the selected patient and each of the patients in the current solution are calculated by the relatedness measure. The similarity is evaluated by the relatedness measure which is calculated by the difference between distances of two patients plus the difference between the time windows of the patients. Then the most similar patient is selected and added to the request bank. The heuristic repeats these steps until $q$ number of patients is placed into the request bank.

**Proximity Removal:** Proximity removal is similar to the Shaw removal heuristic. The proximity removal heuristic considers only the distances between patients when calculating the relatedness measure.

**Time Removal:** Time removal is similar to the Shaw removal heuristic. The time removal heuristic considers only the time windows between patients when calculating the relatedness measure.

**Route Removal:** This heuristic randomly selects a route from the current solution, removes all of the patients from this route, and they are added into the request bank. The idea of the route removal is to redesign the route to minimize the traveled time.

## 3.3    Insertion Heuristics

In general, insertion heuristics are classified as sequential and parallel insertion heuristics in the literature (Solomon, 1987). Sequential insertion heuristics construct the routes one by one. They select a route and try to assign the patients to the selected route without considering the other routes. On the contrary, parallel insertion heuristics consider all of the routes at the same time. Although sequential insertions are faster than parallel insertions, they generally cause slightly worse solutions than parallel heuristics (Liu and Shen, 1999). Therefore, we consider parallel insertion heuristics in this study. Two types of insertion heuristics are used in the proposed algorithms which are *Greedy* and *Regret-k* heuristics. Additionally, we also considered their noise versions because they are shown to improve the solution quality (Pisinger and Ropke, 2007 Hemmelmayr et al., 2012; Riberio and Laporte, 2012; Kovacs et al., 2012).

**Greedy Insertion:** Greedy insertion is a well-known heuristic. All of the patients from the request bank are assigned to all possible positions of the caregivers' routes and an insertion cost is calculated for each position. In this process, only the feasible assignments are considered. After insertion cost is calculated for all patients, the patient with the least insertion cost is assigned to the determined position of the route of the caregiver. This process continues until all patients are assigned to a route or no more insertion is possible. Since only one route of caregivers is changed at each iteration, the insertion cost for other routes does not need to be recalculated at the following iterations. This speed-up idea is applied to all of the insertion heuristics.

**Greedy Insertion with Noise:** The idea of adding noise to the insertion cost is providing randomization to the search process. While the steps of greedy insertion heuristic remain the same, the cost of the insertion is modified by adding the term $t_{max} * \mu * \varepsilon$, where $t_{max}$ is the maximum distance between patients, $\mu$ is the noise parameter that is set to 0.1, and $\varepsilon$ is randomly generated number between [-1, 1].

**Regret-*k* Insertion:** Regret-*k* heuristics are proposed by Potvin and Rousseau (1993). This insertion heuristic improves the greedy insertion. However, it considers not only the patient's best position but also its *k* best positions (depending on choice). Patients are assigned to the positions in order to maximize the regret cost, which is computed as the difference between the costs of *k* best positions. In this respect, the greedy heuristic can be seen as regret-1 heuristic. In our algorithm, we consider regret-*2* and regret-*3* insertions.

**Regret-*k* Insertion with Noise:** This heuristic is a variant of regret-*k* insertion. The insertion cost function is the same as the greedy insertion with noise heuristic.

## 3.4    Lunch Break Position

The abovementioned removal and insertion heuristics only take care of the assignment of the patients to caregivers and their sequence to be visited. Thus, the developed algorithm is enriched to consider lunch breaks by adding the following modification.

While applying removal and insertion heuristics, the break node automatically finds its position in the route. A caregiver has a lunch break when the break node appears before the patient node in its route. Since caregivers can give a break before or after giving service to a patient, the algorithm checks both of the cases for each insertion heuristic phase and accepts the one whose arrival time to the depot is minimized.

## 3.5 Local Search Heuristics

The proposed algorithm contains two local search heuristics to prevent trapping in local optimal solutions. The first one is a well-known Or-opt heuristic which was proposed by Or (1976). The idea of Or-opt is to relocate the selected consecutive patients in a different position without changing the direction of the route. An example of Or-opt could be seen in Figure 1. In this example, nodes $i$ and $i+1$ are selected and their edges $(i-1, i)$ and $(i+1, i+2)$ are deleted (as seen in the left figure). Then the selected nodes are relocated between another selected nodes $j$ and $j+1$ as seen in the right figure.

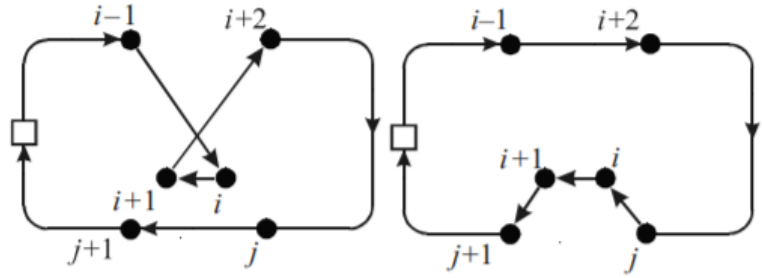

Figure 1. Representation of Or-opt heuristic (sourced from Bräysy and Gendreau (2005).

The second local search heuristic is a novel heuristic that is specifically developed for this problem and called "break local search heuristic". Break local search heuristic simply searches the best lunch break position for the route of each caregiver. The best lunch break position is expected to yield the minimum routing time for caregivers. The pseudocode of the break local search heuristic can be seen in Table 3. Or-opt and break local search heuristics are applied to the new solution at each $\tau_{Or}$ and $\tau_{Break}$ iterations, respectively.

Table 3. The pseudocode of the break local search heuristic algorithm.

**Inputs:** $\pi_k$, route of each caregiver $k$. $a_{i,k}$, arrival time to node $i$ of route $k$.

**for each** $k \in K$
    $\bar{\pi}_k := \pi_k \setminus \{v_b\}$
    **for each** $i \in \pi_k$
        insert lunch break node $v_b$ into before node $i$
        calculate the new route time; $cost_{i,k} = a_{(n+1),k}$
    **end for**
    $i^* = \underset{i}{\operatorname{argmin}}\, cost_{i,k}$
    assign node $v_b$ before node $i^*$ in route $\bar{\pi}_k$
**end for**

Output: $s_{new} = \{\bar{\pi}_1, \dots, \bar{\pi}_m\}$

## 3.6 Update Solution Criteria and Feasibility Check

In order to avoid trapping into local optima, a solution update mechanism is used (Öztürkoğlu and Hoser, 2014; Kocaman et al., 2019; Öztürkoğlu et al., 2019). The solution update subroutine is called if the best solution has not been improved in the last $\omega$ iterations. If this subroutine is called, the best solution is set to a current solution. Random removal is selected as the removal heuristic and Regret-*3* insertion is selected as the insertion heuristic. A new solution is found by applying these heuristics to the current solution. Hence, trapping into a worse local solution could be avoided. A patient can be assigned to a route of a caregiver if and only if the assignment is feasible. In the algorithm, the feasibility check is performed only for time windows constraints.

### 3.7 Acceptance and Stopping Criteria

When a new solution is accepted, it is set as a current solution for the next iteration. For this, we use two policies; (1) a new solution is accepted if its cost is less than the cost of the current solution, (2) otherwise, the new solution still may be accepted with a condition. This probabilistic condition was previously developed for Simulated Annealing (SA) algorithm by Kirkpatrick (1983). Hence, this acceptance criterion also increases the diversification in the algorithm for the global search capability with a probability of $e^{(f(x_{new})-f(x_{curr}))/T}$, where, $f(x_{new})$ and $f(x_{curr})$ are the costs of the new and the current solutions, respectively; $T$ is the temperature, $c$ is the cooling rate that can take a value between 0 and 1. Additionally, the temperature is decreased by multiplying it with a constant cooling rate at each iteration (i.e., $T = T * c$). According to Ropke and Pisinger (2006a, 2006b), the starting temperature $T_{start}$ is found by the initial solution. $T_{start}$ is set to a value that is $\gamma\%$ higher than the value of the initial solution is accepted with probability 0.5 ($T_{start} = -\gamma f(s_{init})/\ln 0.5$).

To determine when to terminate the search process, we first run our algorithm for a predetermined number of iterations $\theta$. When the algorithm reaches $\theta$th iteration, it allows to search $\bar{\theta}$ more number of iterations. If the best solution is improved in the last $\bar{\theta}$ iterations, another $\bar{\theta}$ number of iteration is added to the search process. If the best solution cannot be improved in the last $\bar{\theta}$ number of iterations, the algorithm is terminated and the best solution is reported.

### 3.8 Alternative algorithms developed by the selective local search heuristics

To analyze the effectiveness of the proposed local search heuristics, 4 different variants of the proposed algorithm are considered. The pseudocode of the most comprehensive variant of the proposed ALNS algorithm that contains all of the local search heuristics was previously demonstrated in Table 2. This configuration is named as algorithm A0 hereafter. The second configuration contains only the Or-opt local search heuristic. Therefore, the parameter that controls break local search heuristic, $\tau_{Break}$ is set to a very large number (i.e., $\tau_{Break} = \infty$). This configuration is termed as algorithm A1 hereafter. The third configuration contains the break local search heuristics and called A2 ($\tau_{Or} = \infty$). The last configuration does not contain any local search heuristics ($\tau_{Break}, \tau_{Or} = \infty$) and called algorithm A3 hereafter. It is the base algorithm that provides a comparison between the configurations.

## 4   COMPUTATIONAL STUDY

Liu et al. (2017) used both real data and randomly generated data to test their branch and price algorithm. Although the randomly generated problem instances are publicly available, they did not share the real-life data due to confidentiality. The authors modified classical Solomon (1987)'s VRP instances to generate different HHSRP-LB instances. First, these instances are categorized based on the number of patients and caregivers: 30 patients with 4 caregivers, 50 patients with 5 caregivers, and 100 patients with 12 caregivers. Second, the authors used six classes of Solomon (1987) instances differentiated by the geographical distribution of patients and the tightness of their time windows. Hence, the generated problem instances were defined in a format where the first character shows how patients are distributed in a region, the second character reveal how tight the time windows are, the third and fourth numbers together shows the instance number, and the last numbers after the underline symbol defines the number of patients considered in the instance. For instance, in the problem instance C101_30, "C" identifies that patients are clustered, "1" is for narrower time windows, "01" shows the first instance in this class, and "30" is the number of 30 patients, respectively. Hence, the authors generated 168 problem instances in total. We let readers to review Liu et al. (2017) for the details of the instances.

As described in section 2, the objective function of the problem is to minimize the total traveling cost and total non-visited patient cost. Therefore, a penalty cost for each non-visited patient is 1000 in accordance with Liu et al. (2017). The proposed ALNS algorithm in section 3 was implemented in C# programming language and all experiments are carried out on an Intel Core i7 with 2.20 GHz CPU and 16 GB RAM. Liu et al. (2017) run their branch and price algorithm in a computer with Intel Xeon 2.60 CPU and 128 GB of RAM.

## 4.1 Parameter Tuning

The proposed ALNS algorithm has nine parameters that need to be set before starting the computational study. Because we adopted the general framework of the ALNS algorithm from Ropke and Pisinger (2006a, 2006b), we set the values of the following parameters according to Ropke and Pisinger (2006a, 2006b)'s implementation. Hence, the stopping criterion, which determines when the algorithm stops for searching, has two parameters: (1) the total number of iterations ($\theta$) i.e. 25000, which determines the number of iterations the algorithm runs without any interruption; (2) the additional number of iterations ($\bar{\theta}$) i.e. 1500, which keeps the algorithm running until there is no further improvement in the last $\bar{\theta}$ iterations. Additionally, the cooling rate ($c$), the start temperature control parameter ($\gamma$) and the noise parameters ($r$) were defined as 0.99975, 0.05 and 0.1, respectively. Finally, the Shaw parameters were set as follows: $(\alpha, \beta) = (0.3, 0.1)$.

To determine the best values of the remaining parameters update solution iteration ($\omega$), Or-opt heuristic iteration ($\tau_{Or}$) and break heuristic iteration ($\tau_{Break}$), we conducted a full factorial design of experiment on the algorithm A3 because it consists of all of the three local search heuristics. We determined 5 levels for $\omega$ as $\omega \in \{250, 500, 750, 1000, 1250\}$, and 9 levels for $\tau_{Or}$ ranging from 50 to 250 with a step size of 25. Finally, 9 levels for $\tau_{Break}$ ranging from 75 to 275 with a step size of 25 were specified. We randomly selected 12 instances with 50 patients from Liu et al. (2017) for parameter tuning analysis. Additionally, the algorithm was run for 5 replications using different random number seeds for each instance-setting pair. Hence, we performed 24,300 experiments in total ($9 \times 5 \times 5 \times 5 \times 12$). For an accurate comparison of the different problem instance's objective values, the relative cost percentage ($RCP$) is calculated for each instance-setting pair $i$:

$$RCP_i = \left(\frac{f_i - f_{min}}{f_{min}}\right) * 100$$

where, $f_i$ is the solution found by the algorithm in instance-setting $i$, and $f_{min}$ is the best solution among the experiments for that instance.

We performed ANOVA on Minitab 19 Statistical Software to investigate the effects of parameters on the algorithm's performance with a 95% confidence level. The ANOVA results are given in Table 4. ANOVA results indicate that update solution iteration ($\omega$), break heuristic iteration ($\tau_{Break}$) parameters are statistically significant (their p-values < 0.05) whereas Or-opt heuristic iteration ($\tau_{Or}$) is not significant. In addition, $\omega$ has the greatest effect on RCP value since it has the greatest F-value. Moreover, $\omega \times \tau_{Break}$ is the only significant interaction.

Table 4. ANOVA results for parameter tuning.

| Source | df | Adj. SS | Adj. MS | F-value | P-value |
| --- | --- | --- | --- | --- | --- |
| $\omega$ | 4 | 864.38 | 216.09 | 13.08 | 0.00 |
| $\tau_{Or}$ | 8 | 10.44 | 1.31 | 0.08 | 1.00 |
| $\tau_{Break}$ | 8 | 276.18 | 34.52 | 2.09 | 0.03 |
| $\omega \times \tau_{Or}$ | 32 | 75.46 | 2.36 | 0.14 | 1.00 |
| $\omega \times \tau_{Break}$ | 32 | 3458.17 | 108.07 | 6.54 | 0.00 |
| $\tau_{Or} \times \tau_{Break}$ | 64 | 166.25 | 2.60 | 0.16 | 1.00 |
| $\omega \times \tau_{Or} \times \tau_{Break}$ | 256 | 802.53 | 3.13 | 0.19 | 1.00 |
| Error | 24151 | 395623 | 16.381 | | |
| Total | 24299 | 400473 | | | |

As we discussed in the previous section, we proposed four variants of the ALNS algorithm to investigate whether the implemented local search heuristics have an effect on the quality of the solutions. Whereas algorithms A0, A1, and A2 consist of different or varying numbers of local search heuristics, algorithm A3 does not have any local heuristics but all include $\omega$. Therefore, the parameters were tuned for each algorithm separately according to their availability.

Because algorithm A0 consists of both heuristics, we need to determine the optimal values of each parameter even though the three-way interaction is not statistically significant. Thus, we used the Response Optimizer module of Minitab 19 to determine their optimal values. The module presented that: $(\omega, \tau_{Or}, \tau_{Break}) = (750, 150, 200)$.

In algorithm A1, we only implemented lunch break local search algorithm, we determine the best values of $\omega$ and $\tau_{Break}$. The ANOVA analysis also showed that their interaction is statistically significant. Therefore, using their interaction plot demonstrated in Figure 2, the algorithm presents the lowest cost when $(\omega, \tau_{Break}) = (750, 200)$. Similarly, in algorithm A2, we determine the best parameter values of $\omega$ and $\tau_{Or}$. We also consider their interaction effect plot to determine the best tuning, although their two-way interaction is not statistically significant. As seen in Figure 2, when $(\omega, \tau_{Or}) = (1250, 200)$, the algorithm provides less cost than the other combinations. Since the only $\omega$ is active in algorithm A3, we use its main effect plot to determine its best value. As seen in Figure 3, when $\omega$ is solely set to 1250 it leads the least objective function value. Last, all of the parameters used in the algorithms and their values are summarized in Table 5.

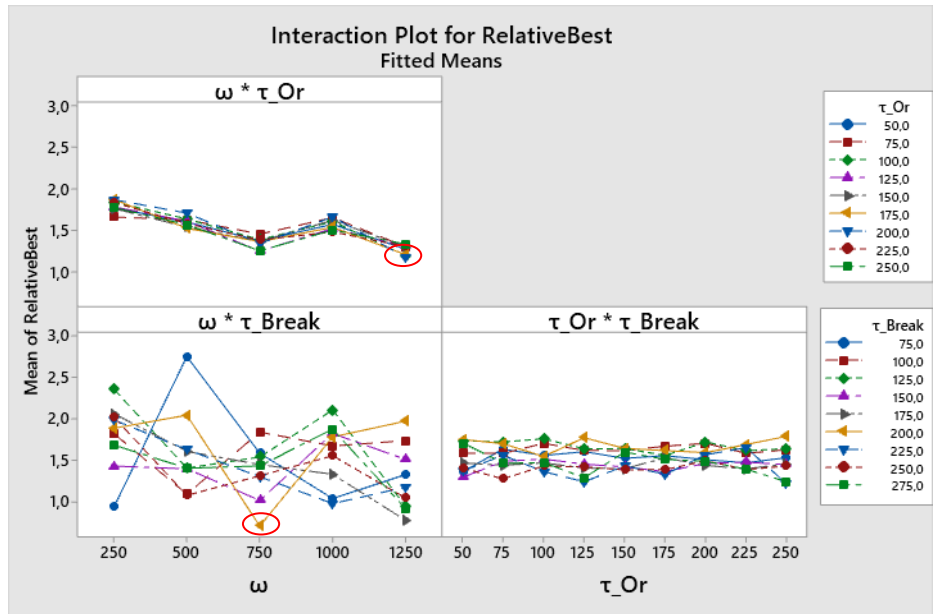

Figure 2. Interaction effects of the parameters.

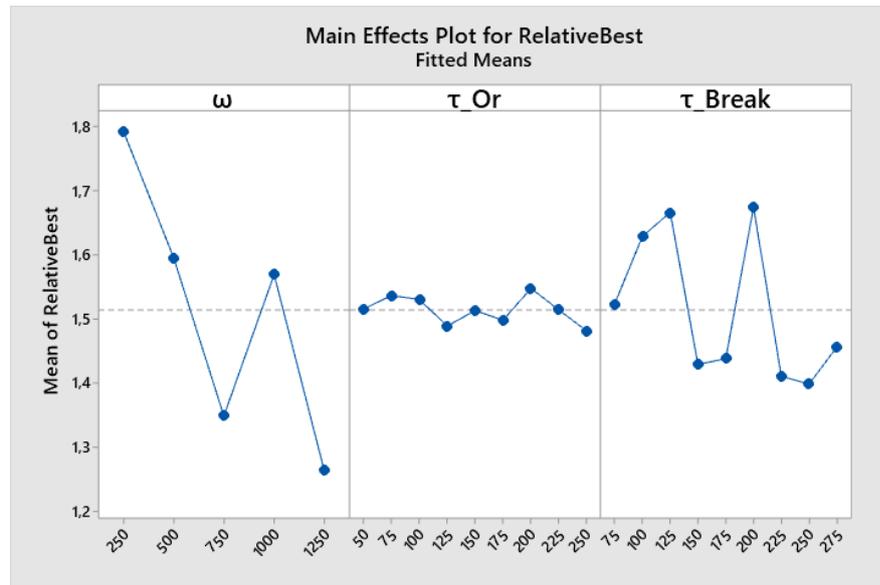

Figure 3. Main effects plot of the parameters.

Table 5. Parameters used in the proposed heuristic.

| Description | Value | Description | Value |
| --- | --- | --- | --- |
| Total number of iterations ($\theta$) | 25,000 | Startup temperature control ($\gamma$) | 0.05 |
| Additional iterations ($\bar{\theta}$) | 1,500 | Cooling rate ($c$) | 0.99975 |
| Update solution iterations ($\omega$) | 750 (in A1 and A0); 1250 (in A2 and A3) | Shaw parameters ($\alpha, \beta$) | (0.3, 0.1) |
| Total number of removed patients | Random (0.1n, 0.3n) | Noise ($r$) | 0.1 |
| Or-opt heuristic iteration ($\tau_{Or}$) | 150 (in A0); 200 (in A2) | Break heuristic iteration ($\tau_{Break}$) | 200 (in A0 and A1) |

## 4.2 Computational Results

For each problem instance, we executed 5 replications to test the robustness of the proposed algorithms in terms of the best-found solutions so far. Thus, we first compare the effectiveness of the proposed algorithms in terms of their best-found solutions to investigate the effect of proposed local search heuristics. We then also compare their computational times to see whether the implementation of local search heuristics suffers a large computational burden. Last, we compare the performance of the proposed algorithms with Liu et al. (2017)'s best-found solutions.

*4.2.1 Comparisons of the proposed algorithms in terms of their best-found solutions*

Tables A1, A2 and A3 in the appendix present the best of best-found solutions (called *Best* hereafter) by algorithms A0, A1, A2 and A3 within five replications for 30-, 50- and 100-patients instances. Additionally, the number of unvisited patients in the *Best* and the average computational time of the five replications were presented in these tables. For the sake of clarity, the average of the best solutions was also presented only for algorithm A0.

We conducted a one-way ANOVA using the *Best* solutions obtained by each algorithm to test whether their means of the best objectives are the same or not. Thus, the null hypothesis $H_0$ is designed as the means of the algorithms' best objective values are the same. Using Minitab 19 Statistical Software with 95% confidence interval, the demonstrated results in Table 6 showed there is no statistical difference between the algorithms in terms of their objective values regardless of the number of patients considered in the problem instances ($H_0$ is rejected because the p-value is 1.0). Moreover, Tukey's test of pairwise comparison also shows that the proposed algorithms could be grouped in the same class.

Table 6. The comparison of the proposed algorithms in terms of their objective values.

| Instance Groups | One-way ANOVA results | | | | | | Grouping information using Tukey Method | | | |
|---|---|---|---|---|---|---|---|---|---|---|
| | Source | DF | SS | MS | F | p | Rank | N | Mean | Grouping |
| 30-patient instances | Factor | 3 | 16 | 5 | 0.00 | 1.0 | A1 | 56 | 1855 | A |
| | Error | 220 | 1325624854 | 6025568 | | | A0 | 56 | 1854 | A |
| | Total | 223 | 1325624870 | | | | A3 | 56 | 1854 | A |
| | S = 2455   R-Sq = 0.00%   R-Sq(adj) = 0.00% | | | | | | A2 | 56 | 1854 | A |
| 50-patient instances | Factor | 3 | 94523 | 31508 | 0.00 | 1.0 | A2 | 56 | 5210 | A |
| | Error | 220 | 8104196492 | 36837257 | | | A3 | 56 | 5209 | A |
| | Total | 223 | 8104291015 | | | | A0 | 56 | 5178 | A |
| | S = 6069   R-Sq = 0.00%   R-Sq(adj) = 0.00% | | | | | | A1 | 56 | 5162 | A |
| 100-patient instances | Factor | 3 | 142549 | 47516 | 0.00 | 1.0 | A3 | 56 | 10153 | A |
| | Error | 220 | 25590578913 | 116320813 | | | A1 | 56 | 10141 | A |
| | Total | 223 | 25590721462 | | | | A2 | 56 | 10136 | A |
| | S = 10785   R-Sq = 0.00%   R-Sq(adj) = 0.00% | | | | | | A0 | 56 | 10087 | A |

Even though the variants are not statistically different, the algorithm A0 is slightly diverged from the others by providing slightly lower costs large problem instances (A0 provided the lowest mean as 10087). Thus, we can conclude that the insertion and the removal heuristics used in the ALNS algorithm with the selected update parameters seem to work efficiently for large instances.

We also should highlight that considering the lunch break position (discussed in section 3.4) during removal and insertion processes seem to reduce the need for an additional local search heuristic for positioning lunch break.

*4.2.2 Comparisons of the proposed algorithms in terms of their average solution times*

Similar to the previous section, we performed a one-way ANOVA for investigating the effect of local search heuristics on the mean computational times of the algorithms. Table 7 presents the ANOVA results with grouping information obtained by Tukey's pairwise test. For problem instances that are generated by 30 and 50 patients, the null hypothesis, where the mean computational time of each algorithm is the same, is rejected due to p=0.00 with a 95% confidence. However, the Tukey's pairwise comparison test shows that while algorithms A0, A1 and A2 take statistically the same amount of time the algorithm A3 diverges from them with lesser run time. Thus, this result shows that the implementation of the local search heuristics requires extra time but not significantly different from each other to find the best solutions although they do not provide a statistically significant improvement on the quality of the solutions according to the discussions in the previous section.

When the problem size increases to 100 patients, it was seen that there is no statistical difference between the algorithms in terms of run time (see Table 7). Thus, this result shows that the algorithm A3 needs to run more time to find its best solutions than before because it does not utilize any local search to improve the current solution.

Table 7. The comparison of the proposed algorithms in terms of their computational time.

| Instance Groups | One-way ANOVA results | | | | | | Grouping information using Tukey Method | | | |
|---|---|---|---|---|---|---|---|---|---|---|
| | Source | DF | SS | MS | F | p | Rank | N | Mean | Grouping |
| **30-patient instances** | Factor | 3 | 655.05 | 218.35 | 26.90 | 0.00 | A1 | 56 | 13.27 | A |
| | Error | 220 | 1785.93 | 8.12 | | | A2 | 56 | 12.45 | A |
| | Total | 223 | 2440.98 | | | | A0 | 56 | 12.41 | A |
| | S = 2.849   R-Sq = 26.84%   R-Sq(adj) = 25.84% | | | | | | A3 | 56 | 8.84 | B |
| **50-patient instances** | Factor | 3 | 1449.33 | 483.11 | 50.03 | 0.00 | A1 | 56 | 23.29 | A |
| | Error | 220 | 2124.63 | 9.66 | | | A0 | 56 | 23.29 | A |
| | Total | 223 | 3573.96 | | | | A2 | 56 | 22.32 | A |
| | S = 3.108   R-Sq = 40.55%   R-Sq(adj) = 39.74% | | | | | | A3 | 56 | 17.16 | B |
| **100-patient instances** | Factor | 3 | 140.1 | 46.7 | 1.13 | 0.34 | A3 | 56 | 54.88 | A |
| | Error | 220 | 9091.4 | 41.3 | | | A0 | 56 | 54.05 | A |
| | Total | 223 | 9231.5 | | | | A2 | 56 | 53.16 | A |
| | S = 6.428   R-Sq = 1.52%   R-Sq(adj) = 0.17% | | | | | | A1 | 56 | 52.86 | A |

*4.2.3 Comparisons of the proposed algorithms with Liu et al. (2017)*

The best and the average solutions of the five replications are presented in Tables A1, A2 and A3 in the appendix in detail. In Table 8, the best-found solutions of the algorithms, called *Alg_Best* hereafter, are compared with Liu et al. (2017)'s best solutions, called *Liu_Best* hereafter, briefly. Because Liu et al. (2017) could not provide either a feasible or an optimal solution for some problem instances, we listed the number of optimal solutions found by our algorithms as *#Opt* while *#LiuOpt* is the number of known optimal solutions provided by Liu et al. (2017). Similarly, *#Imp* and *#Worse* indicate the number of improved and worse *Alg_Best* in comparison with *Liu_Best*. *AvgGap* is the percentage of the gap between *Alg_Best* and *Liu_Best*, and calculated as 100x(*Liu_Best* - *Alg_Best*)/ *Liu_Best*, where the positive value indicates an improvement over *Liu_Best*. Similarly, *StdGap* indicates the standard deviation of the *AvgGap*. Last, *AvgCpu* and *AvgCpu_Liu* indicate the average computational time of the algorithms to solve instances in seconds.

As seen in Table 8, the algorithm A0 slightly outperforms the other algorithms based on *#Opt*, *#Imp*, and *AvgGap*. The algorithm improved 36 solutions and found 88% of the optimal solutions. According to *AvgGap*, the worse solutions seem very close to *Liu_Best* of which they are only 0.03% and 0.02% away in 30- and 50-patient instances on average, respectively. Moreover, the algorithm presents about 0.7% improvement in 100-patient problem instances. The other algorithms also obtained the majority of the optimal solutions, 83% on average, and provided lower costs than *Liu_Best* in 36 of the total instances on average.

Table 8. The brief results of the comparisons of the proposed algorithms with Liu et al. (2017)' best solutions.

| Proposed algorithm | Instances with #patients | #LiuOpt | #Opt | #Imp | #Worse | AvgGap (%) | StdGap | AvgCpu (s) | AvgCpu_Liu (s) |
|---|---|---|---|---|---|---|---|---|---|
| A0 | 30 | 54 | 53 | 1 | 2 | -0.03 | 0.17 | 12.41 | 1106.3 |
|  | 50 | 44 | 38 | 11 | 7 | -0.02 | 1.13 | 23.29 | 1075.5 |
|  | 100 | 22 | 14 | 24 | 18 | 0.66 | 3.23 | 54.05 | 2509.7 |
| A1 | 30 | 54 | 52 | 1 | 3 | -0.01 | 0.07 | 13.27 | 1106.3 |
|  | 50 | 44 | 33 | 11 | 12 | -0.28 | 1.76 | 23.29 | 1075.5 |
|  | 100 | 22 | 12 | 25 | 19 | 0.15 | 3.57 | 52.86 | 2509.7 |
| A2 | 30 | 54 | 52 | 1 | 3 | -0.04 | 0.21 | 12.45 | 1106.3 |
|  | 50 | 44 | 37 | 11 | 8 | 0.09 | 0.74 | 22.32 | 1075.5 |
|  | 100 | 22 | 12 | 22 | 22 | -0.10 | 3.97 | 53.16 | 2509.7 |
| A3 | 30 | 54 | 52 | 1 | 3 | -0.01 | 0.07 | 8.84 | 1106.3 |
|  | 50 | 44 | 35 | 11 | 10 | -0.26 | 1.75 | 17.16 | 1075.5 |
|  | 100 | 22 | 12 | 24 | 20 | -0.23 | 4.98 | 54.88 | 2509.7 |

Because the algorithm A0 provides better solutions, even if slightly than the other algorithms, we selected this for comprehensive comparison analysis with the *Liu_Best*. Table A4 in the appendix presents the results of this comparison in detail. Furthermore, Table 9 briefly provides some information about the data in Table A4. The notations in Table 9 are explained in the following paragraphs. Some of the notations used in Table A4 are also explained in the following for clarification, the others are explained in the appendix.

In Table A4 in the appendix, *OptValAvg* and *OptValStd* refer the mean and the standard deviations of the *Liu_Best*s in the given set of problem instances. Similarly, *Avg.Unvisited* indicate the average number of unvisited patients in *Liu_Best*s for the specified instance class. To make an accurate comparison, *BestAvg* and *BestStd* define the mean and the standard deviations of the *Best* of the algorithm A0. *Avg.BestUnvisited* indicates the average number of unvisited patients in *Best* of the algorithm A0.

In Table 9, *Dif_BestAvg* is the percentage of the difference between *OptVal Avg* and *BestAvg* and calculated as *100(OptValAvg – BestAvg)/OptValAvg*. *CV_Best* and *CV_Liu* refer to the coefficient of variance of the *Best* and *Liu_Best*. *Dif_AvgUnvisited* is equal to (*Avg.Unvisited – Avg.BestUnvisited* ). The following observations could be made from the results in Table 9 and Table A4.

1) While Liu et al. (2017) presented 120 optimal solutions out of 168 total instances the algorithm A0 obtained 105 of the optimal solutions and provided 36 better solutions than that. Especially for large instances with 100 patients, the algorithm A0 improved solutions in 24 instances where Liu et al. (2017) presented solutions with an average gap of 8.3% because of the complexity of these instances. As the number of patients increases, the complexity of the HHSRP-LB is expected to be increased. Therefore, Liu et al. (2017) presented solutions with an average gap of 5% and 8.3% for 50- and 100-patient instances. Especially when patients are randomly distributed and they have narrow time windows (class R1) the average gap increased to 14% and 22%. The algorithm A0 improved the quality of the solutions on average 6% and 16% in 50- and 100-patient instances, respectively. Especially for R1 instances with 50 and 100 patients respectively, the algorithm presented 18.4% and 35% lower costs than *LiuBest* on average.

2) As known that the model penalizes if there are unvisited patients due to the working hour constraint. Hence, this penalty aims to cover as many patients as possible to increase home care service coverage and patients' satisfaction. As seen in column *Dif_AvgUnvisited* in Table Table 9, the algorithm A0 achieved to visit more patients than Liu et al. (2017)'s algorithm on average especially for large instances with large gaps. For instance, the algorithm accomplished to visit about 10 and 6 more patients in instance classes R1 and RC1 with 100 patients, respectively, and 2 more patients in R1 with 50 patients.

3) Similar to the Liu et al. (2017)'s algorithm, the algorithm A0 could also be seen as robust algorithm because the average coefficient of variance of the *Best* solutions is 0.3 on average as *LiuBest* has 0.27.

4) Last, the algorithm A0 powered by two additional local heuristics require very less amount of time to generate its solutions than Liu et al. (2017) as expected. On average, while *LiuBest* required about 1100, 1075 and 2510 seconds, the proposed algorithm runs about 12, 23 and 54 seconds for solving 30-, 50- and 100-patient instances. We would like to note that Liu et al. (2017) run their model in a much more powerful computer than ours and they couldn't obtain any solution even after running three hours. Additionally, the computational burden of the proposed algorithm does not increase significantly, which seems to increase linearly, as the size of the instances increases.

In a conclusion, the prosed algorithm is effective for use in practice for a daily home health care scheduling and routing planning because of the quality of the solutions, its robustness and a low computational requirement. Moreover, because the proposed ALNS algorithm is less complex than Liu et al. (2017)'s branch-and-price algorithm, it could be easily embedded in a decision support system for an efficient planning in practice.

Table 9. The comparisons of *Best* by algorithm A0 and *Liu_Best*.

| Instance Classes | *Dif_BestAvg (%)* | *CV_Best* | *CV_Liu* | *Dif_AvgUnvisited* | *#Opt* | *#Imp* | *#Worse* |
|---|---|---|---|---|---|---|---|
| C1_30 | 0.00 | 0.79 | 0.79 | 0.00 | 9 | 0 | 0 |
| C2_30 | 0.00 | 0.02 | 0.02 | 0.00 | 8 | 0 | 0 |
| R1_30 | 0.00 | 0.83 | 0.83 | 0.00 | 12 | 0 | 0 |
| R2_30 | 0.01 | 0.06 | 0.06 | 0.00 | 11 | 0 | 0 |
| RC1_30 | 4.00 | 0.31 | 0.29 | 0.2 | 7 | 0 | 1 |
| RC2_30 | 1.92 | 0.11 | 0.10 | 0.00 | 6 | 1 | 1 |
| C1_50 | 8.77 | 0.92 | 0.85 | 0.2 | 8 | 1 | 0 |
| C2_50 | 1.41 | 0.05 | 0.04 | 0.00 | 6 | 1 | 1 |
| R1_50 | 18.41 | 0.49 | 0.36 | 2.3 | 6 | 4 | 2 |
| R2_50 | 5.64 | 0.10 | 0.08 | 0.00 | 6 | 3 | 2 |
| RC1_50 | -0.80 | 0.24 | 0.23 | -0.1 | 7 | 0 | 1 |
| RC2_50 | 2.55 | 0.10 | 0.10 | 0.00 | 5 | 2 | 1 |
| C1_100 | 16.38 | 0.63 | 0.53 | 1.7 | 1 | 4 | 4 |
| C2_100 | 1.68 | 0.07 | 0.05 | 0.00 | 5 | 0 | 3 |
| R1_100 | 35.15 | 0.49 | 0.28 | 10.4 | 1 | 5 | 6 |
| R2_100 | 16.65 | 0.08 | 0.00 | 0.00 | 1 | 10 | 0 |
| RC1_100 | 19.86 | 0.24 | 0.12 | 6.4 | 3 | 0 | 5 |
| RC2_100 | 7.71 | 0.09 | 0.06 | 0.00 | 3 | 5 | 0 |

## 5 HOME HEALTHCARE DECISION SUPPORT SYSTEM (HHCSS)

Decision Support System (DSS) is an information system that permits users to seek help from computer technology during decision making. It is a combination of data, information, software, analysis, and mathematical model which helps people to understand the complex systems and solution methodology of these systems. With an aim of assisting experts (end-users) in their decision-making and the families of the patients to have better service for the health care of their loved once, a prototype home healthcare decision support system (HHCSS) has also been developed for the HHC problem under study. For implementing the desktop application of the proposed system, different libraries such as matplotlib[1], NumPy[2], scikit-learn[3], and Tkinter[4] have been used, in which the Tkinter in the standard GUI library for python. Tkinter provides a powerful object-oriented environment to the GUI toolkit. The integration of python and Tkinter allows users to create a GUI application through a fast and easy process.

### 5.1 Architecture of HHCSS

The designed HHCSS is a model-driven single installation system that has all the essentials programs and databases stored locally. The HHCSS architect is divided into three main parts. "Data Entry Form", "Solver" and "Visualization" as shown in Figure 4. Through the "Entry Forms", a new patient's, caregivers, and vehicle data will be entered in the database of the system which will be saved in .xlsx format. Other than that the information about the existing patients, caregivers, and vehicles can also be seen at the bottom of their respective menus as illustrated in Figure 5.

---
[1] Matplotlib initially developed by Hunter (2007) is a comprehensive library for creating static, animated, and interactive visualizations in Python. It can be freely accessed from https://matplotlib.org.
[2] Numpy is a 100% open scientific computing software with Python which is accessible from https://numpy.org.
[3] Sklearn or scikit-learn is an open machine learning sofrware with Python which was initially developed by Pedregosa et al. (2011) and is accessible from https://scikit-learn.org/stable/
[4] Lundh, F. (1999). An introduction to tkinter. URL: www.pythonware.com/library/tkinter/introduction/index.htm

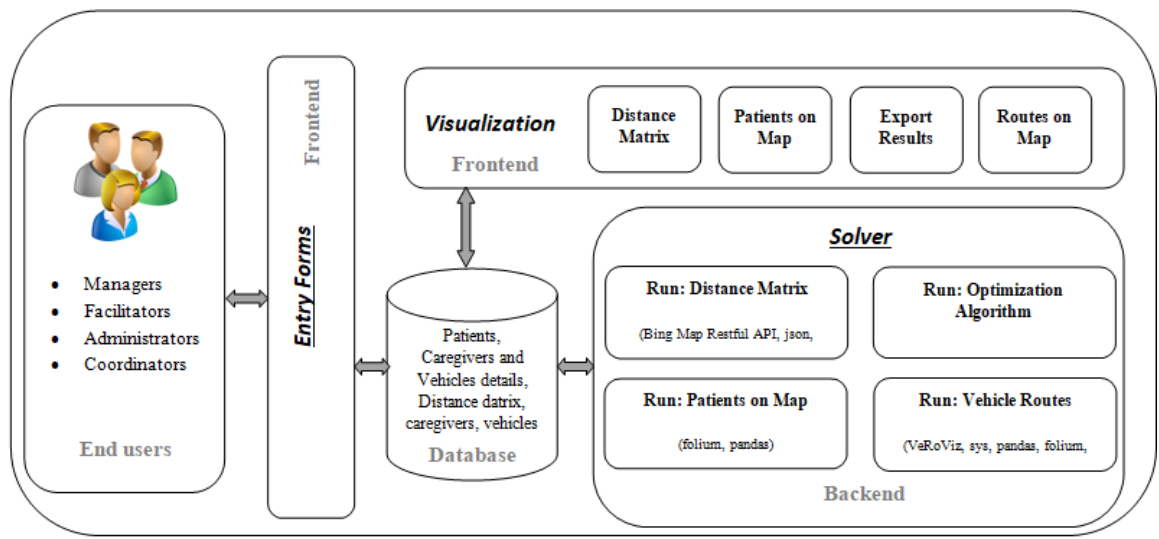

Figure 4: Architecture of HHCSS

"Solver" and "Visualization" parts can be seen in the extreme left of the HHCSS under the menu option, as shown in Figure 6. "Solver" is further divided into four different modules which are distance matrix generator, patients' location on the map, optimization algorithm, and routes of the vehicle on a map. These modules can be operated by the respective buttons which will exploit the required data available in the database for the generation of the desired results. These buttons include "Run: Distance Matrix", "Run: Patients on Map", "Run: Optimization Algorithm" and "Run: Vehicles Routes". In short, all the buttons starting with "Run:" are the part of "Solver" and will use to run the code for the desired operation.

The data which is in the database or entered through the "Patients Entry Form" will be used for calculating the road distances from every node to every node by clicking the "Run: Distance Matrix" button. The actual road distance matrix will be obtained by using Bing$^{TM}$ Maps Restful API[5] for which a python code is developed using JSON[6] and urllib[7] libraries. The Bing$^{TM}$ map distance matrix API[8] provides travel time and distances for a set of origins and destinations. The distances and times returned are based on the routes, calculated by the bing maps route API. The distance matrix can be estimated for three different types of routes, including driving, walking, and public transit. In HHCSS, it is assumed that all distances between patients and the HHC are covered by typical cars. The generated distance matrix can be seen in .xlsx and .txt format by clicking the "Distance Matrix" button from the menu.

---

[5] The Bing$^{TM}$ Maps Restful API provides interface to perform tasks regarding maps geocoding, creating routes, etc. https://docs.microsoft.com.
[6] Pezoa, F., Reutter, J. L., Suarez, F., Ugarte, M., & Vrgoč, D. (2016, April). Foundations of JSON schema. In *Proceedings of the 25th International Conference on World Wide Web* (pp. 263-273).
[7] urllib is a python package for dealing with URLs. It can freely accesed from https://docs.python.org/2/library/urllib.html
[8] The Bing$^{TM}$ Maps Distance Matrix API helps in determining the best routes. It can freely accessed for limited nodes from https://www.microsoft.com/en-us/maps/distance-matrix

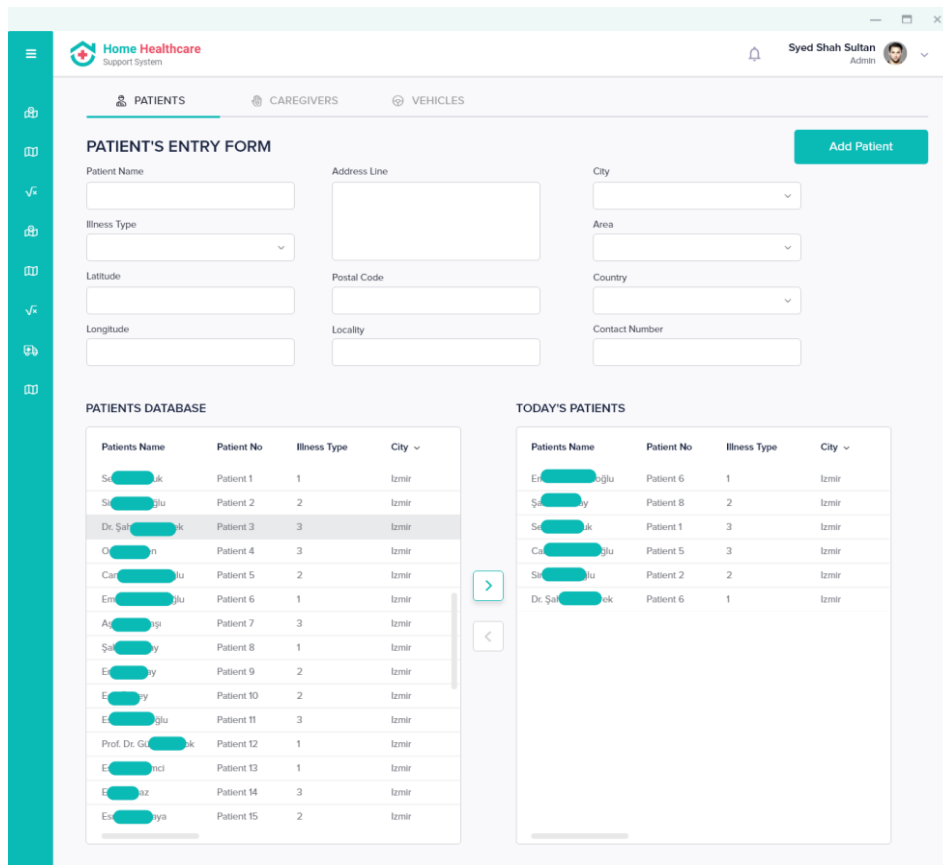

Figure 5. Outlook of HHCSS Outlook of HHCSS with the menu

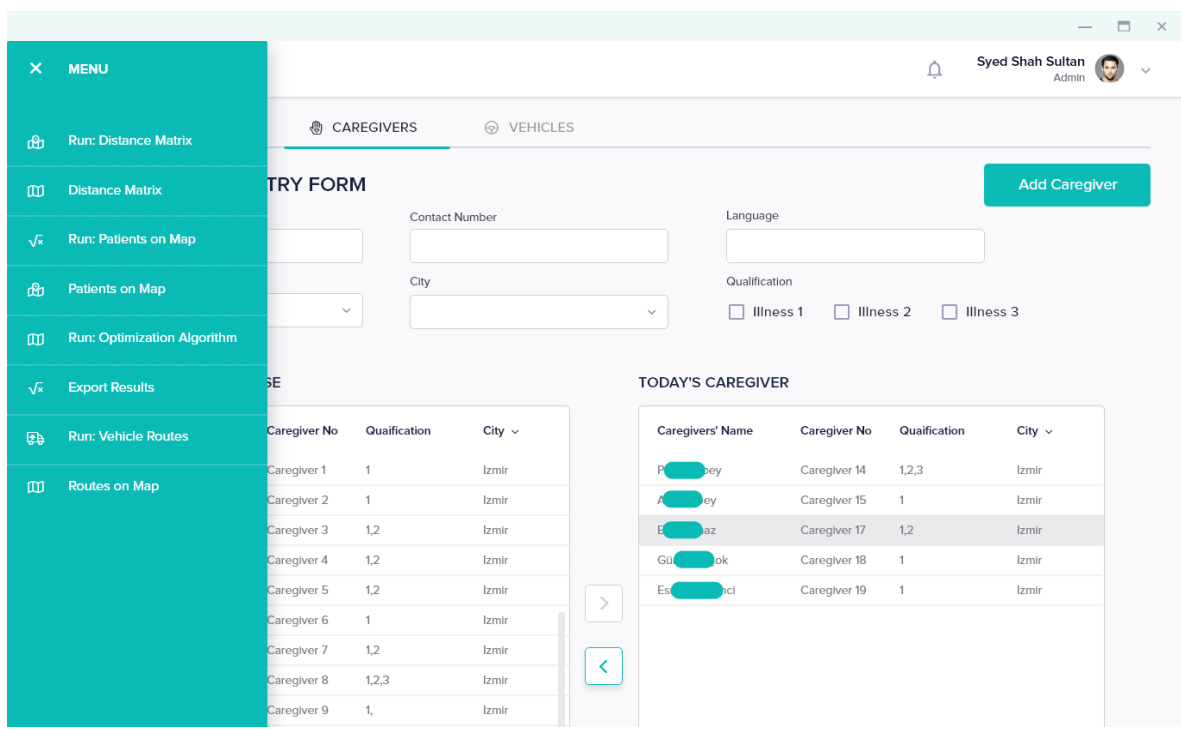

Figure 6. Outlook of HHCSS with the menu

To get a better idea of patients' locations, a code is written by using the folium[9] library and can be run by clicking the "Run: Patients on Map" button. Folium library is very helpful in visualizing the data that's been manipulated in Python on an interactive Leaflet map[10]. For the input, the data that was used for calculating the distance matrix is utilized here also through the panda's library. The results of that code can be seen as an HTML file by clicking the next button of the pair i.e."Patients Map".

For optimizing the vehicle routes, the ALNS code (Algorithm A0) is used which has been explained in section 3. The ALNS is taking three inputs that are distance matrix, patients data, and the caregivers' information. By clicking the "Run: Optimization Algorithm" .exe file will be executed from the python environment for generating the result. The result of vehicle route optimization can be seen through the "Export Results" button.

The code for the shortest vehicles' route according to the actual road network is written in Python 3 and can be run by clicking "Run: Vehicle Routes". VeRoViz (Vehicle Routing Visualization) (Peng and Murray, 2019), an open-source package for generating and visualizing the nodes and vehicle routes on the road networks and an API from Open Route Service (ORS)[11], a directional service is used while coding. The result of this coding can be seen by clicking the last remaining button i.e."Road Routes on Map".

## 5.2    Visualization

As far as the "Visualization" part is concerned, all the remaining buttons in the menu that are "Distance Matrix", "Patients on Map", "Export Results" and "Routes on Map" are incorporated in it. These buttons will display the results of the code that was called with the previous buttons in the pair. Figure 7 will give a better idea about the whole process of HHCSS.

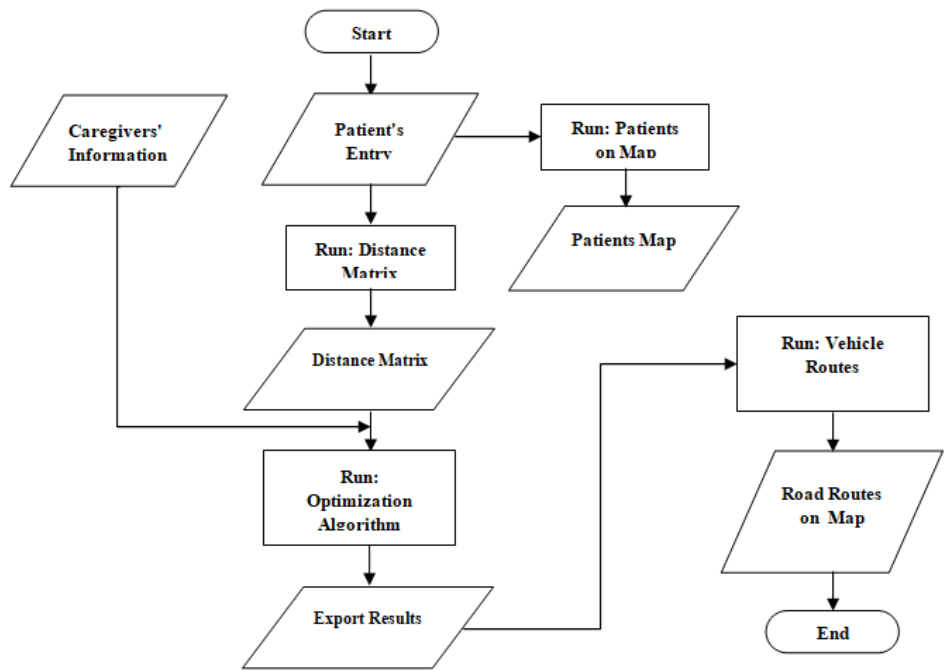

Figure 7: Flowchart of HHCSS

---

As it is already mentioned above that the road distance matrix of the respective patients' size has been generated through Bing Map API using the coordinates of all the patients including a hospital. This distance matrix is further be inserted as an input in algorithm A0 that is the best variant of the proposed ALNS Lunch Break algorithm, explained in section 3.8. The result of the model is used to see the sequence of visiting the patients and road routes of all the vehicles on the map.

Figure 8 shows an example visualization of the locations of patients on a map that need to be visited with a distinct color variation depending on the area in Izmir, Turkey. This map is obtained by running the backend code of the "Run Code: Map on Patients" button.

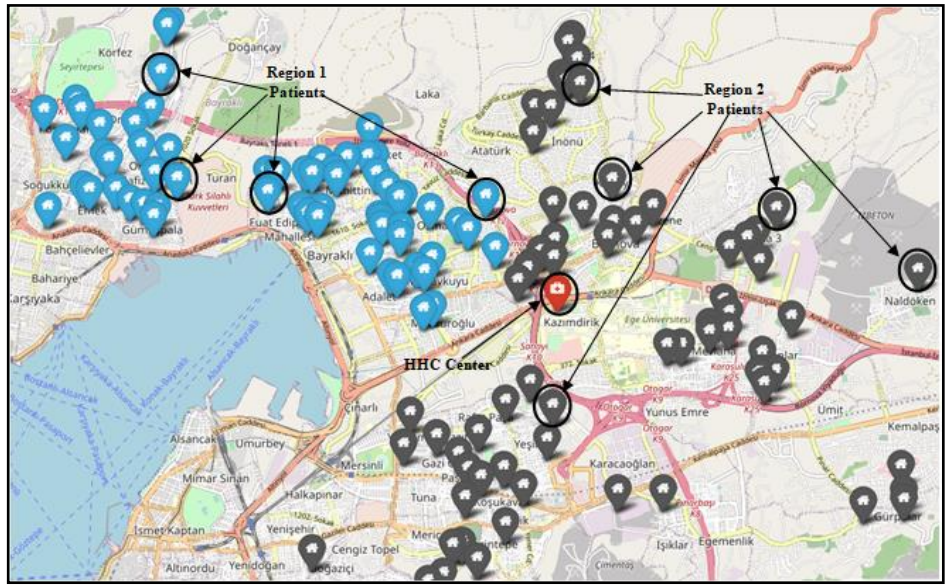

*Region 1: Bayrakli (Izmir); Region 2: Bornova (Izmir)

Figure 8: Example representation of patients' locations on the map in Izmir, Turkey.

## 6   EXPERIMENTS AND EMPIRICAL RESULTS ON THE DATA SET OF COVID-19 PATIENTS

In the following section, the proposed algorithm will be analyzed on the data of the COVID-19 patients and the results obtained from it will be studied. For this, the approximate locations of COVID-19 patients present in two neighboring districts of the three biggest cities of Turkey, namely Ankara, Istanbul, and Izmir have been extracted. These three cities are at high risk of spreading the virus that causes COVID-19 to people (COVID-19 Istanbul, Ankara ve Izmir density and risk map, September 07, 2020) and on average 200-250 positive cases are being observed per day just in Izmir (Izmir Coronavirus table, 2020). Neighboring districts were selected from these cities in terms of importance in diplomatic affairs, tourism, and their role as financial centers. This extraction of the approximate locations from these districts were made through the heatmap available in the mobile application "Life Fits into Home" (TR Ministry of Health, 2020). This mobile application was developed and continuously updated by the digital transformation office of the Turkish government. The main objective behind this application is to protect its citizens from being exposed to dangerous areas during the current pandemic. Figure 9 provides heat maps of COVID-19 patients from various provinces in Turkey that was extracted from the application.

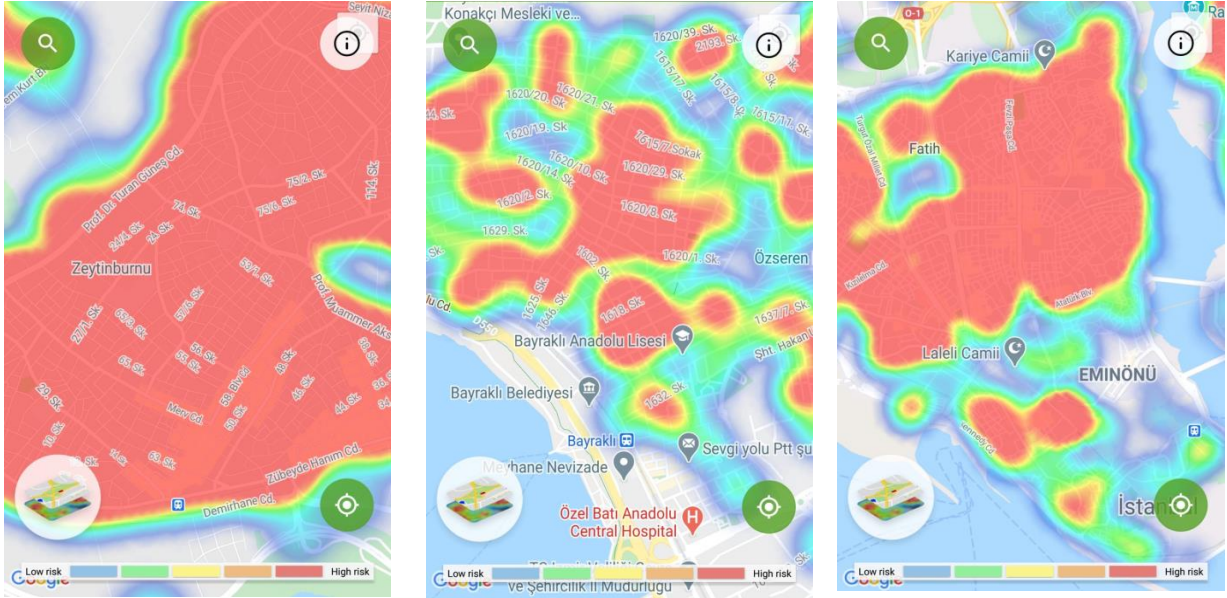

Figure 9: Heat map images from "Life Fits into Home" application

### 6.1 Test Instances

We generated small to large instances from the approximate patients' lists in the selected districts in Istanbul, Ankara and Izmir in Turkey. The largest government hospital, located within the neighborhood of the selected districts has been assigned as the home healthcare center to provide enough caregivers to serve all the patients. Details of the selected districts in the three biggest cities of Turkey and the number of patients concerning cities along with their selected districts are given in Table 10. The symbols for the cities, which are further be used to represent instances are also in brackets with their names. The minimum, average, and maximum distance of patients from the HHC center is also given on the right-hand side of the same table.

Table 10. Details of the selected neighborhood with respect to their cities, the number of caregivers.

| City | Neighboring Districts | # Patients | Land Areas (km$^2$) | Total # Patients | Distance to HHC (km) | | |
|---|---|---|---|---|---|---|---|
| | | | | | Min. | Avg. | Max. |
| Ankara (AN) | Çankaya (C) | 195 | 268 | 344 | 0.68 | 12.6 | 24 |
| | Altındağ (A) | 149 | 174.5 | | | | |
| Istanbul (IS) | Fatih (F) | 101 | 13.08 | 184 | 0.78 | 4.78 | 11.61 |
| | Zeytinburnu (Z) | 83 | 12.08 | | | | |
| Izmir (IZ) | Bornova (O) | 68 | 224 | 131 | 1.44 | 6.56 | 12.71 |
| | Bayraklı (B) | 63 | 30 | | | | |

In total there are six classes of instances generated for each city that are given in Table 11. These instances are going to be represented in the format like 50SO1, 100IS2, ANAC344 which means 50 randomly selected patients from Bornova in the first instance, 100 randomly selected patients from whole Istanbul out of 184 patients in total in the second instance, and all patients of Ankara (A+C) that are 344 respectively.

Table 11. Details of the different classes of instances within cities

| Classes of Instances (Ankara) | Classes of Instances (Istanbul) | Classes of Instances (Izmir) |
|---|---|---|
| SA: Sample of Altindag | SF: Sample of Fatih | SO: sample of Bornova |
| SC: Sample of Cankaya | SZ: Sample of Zeytinburnu | SB: sample of Bayrakli |
| AN: Sample of Ankara | IS: Sample of Istanbul | IZ: Sample of Izmir (B+O) |
| ANA: Whole Altindag | ISF: Whole Fatih | IZB: Whole Bayrakli |
| ANC: Whole Cankaya | ISZ: Whole Zeytinburnu | IZO: Whole Bornova |
| ANAC: Whole Ankara (A+C) | ISFZ: Whole Istanbul (F+Z) | IZBO: Whole Izmir (B+O) |

Three different types of services for the caregivers have been considered based on our experience and interviews with practitioners. Type-I service is assumed to be for diagnosing the COVID-19 test such as PCR test. In this type of service, a caregiver visits the patient to take samples for diagnosing the disease. Type-II service is assumed to be for the simple medication and varing of positive COVID-19 patients who does not have a chronic disease, whereas the Type-III service includes the medication of positive COVID-19 patients and their chronic diseases but not in a high-risk group. Patients who are in the high-risk group are not included because they have to be treated in hospitals or in some specially built isolation centers. Therefore, the proposed service types are a standard procedure with low variability. However, the service times of different types of services differ from one another. The mean and coefficient of variation values are estimated for each type of service. The mean service time values of Type-I, Type-II, and Type-III services are assumed to be 10, 15, and 20 minutes, respectively. The coefficient of variation of all types of services is estimated as 0.25 due to low variability caused by cultural behaviors. In addition, after talking to practitioners, we assumed that 60%, 30%, and 10% of patients seek Type-I, Type-II, and Type-III treatments, respectively. On the other hand, hierarchical qualification levels are determined for caregivers which means all of the caregivers can treat Type-I services, 50% of caregivers can treat Type-II, and only 20% of caregivers can treat Type-III services. Lastly, caregivers work between 9:00 and 18:00 and must have 60 minutes lunch break between 11:00 and 14:00. The details of the generated patients' and caregivers' data concerning types of services are given in Table 12 in which the number of caregivers has been defined with the assumption to represent practical concerns.

Table 12. Patients' and caregivers' detail concerning the type of services

| # patients | # available caregivers | # patients suffering according to the type of illnesses. | | | # available caregivers with respect to the type of illnesses. | | |
|---|---|---|---|---|---|---|---|
| | | Type-I | Type-II | Type-III | Type-I | Type-II | Type-III |
| 30 | 2 | 18 | 9 | 3 | 2 | 1 | 0 |
| 50 | 3 | 30 | 15 | 5 | 3 | 2 | 1 |
| 100 | 5 | 60 | 30 | 10 | 5 | 3 | 1 |
| 63 (Bayrakli) | 4 | 38 | 19 | 6 | 4 | 2 | 1 |
| 68 (Bornova) | 4 | 41 | 20 | 7 | 4 | 2 | 1 |
| 131 (Izmir) | 7 | 79 | 39 | 13 | 7 | 4 | 1 |
| 40 | 2 | 24 | 12 | 4 | 2 | 1 | 0 |
| 60 | 3 | 36 | 18 | 6 | 3 | 2 | 1 |
| 100 | 5 | 60 | 30 | 10 | 5 | 3 | 1 |
| 101 (Fatih) | 6 | 61 | 30 | 10 | 6 | 3 | 1 |
| 83 (Zeytinburnu) | 5 | 50 | 25 | 8 | 5 | 3 | 1 |
| 184 (Istanbul) | 10 | 110 | 55 | 19 | 10 | 5 | 2 |
| 75 | 4 | 45 | 23 | 7 | 4 | 2 | 1 |
| 100 | 5 | 54 | 27 | 9 | 5 | 3 | 1 |
| 149 (Altindag) | 8 | 89 | 45 | 15 | 8 | 4 | 2 |
| 195 (Cankaya) | 10 | 117 | 59 | 19 | 10 | 5 | 2 |
| 344 (Ankara) | 18 | 206 | 104 | 34 | 18 | 9 | 4 |

## 6.2 Experiment and results

In this section, the results from various approximately generated patients' instances of COVID-19 have been discussed. Tables A5-A7 in Appendix show the details of the results of the generated problem instances of COVID-19 patients in each city. For the sake of clarification, Table 13 demonstrates the summary of those results that are the averages of the corresponding test instances of COVID-19 patients' data concerning the city and patients' size. These results are obtained using the developed DSS. The average of the total travel time spent by caregivers for the all corresponding instances is identified by "Avg. Total Travel Time" in minutes whereas "Avg. Total Working Time" shows how much time all caregivers spent for servicing and traveling in a single day on average. For an instance, two caregivers spent 44.46 minutes for traveling and 437.46 minutes (implicity 393 min. for caring) for both servicing and travelling all 30 patients in Bayrakli on average. Hence, average total service time can be calculated by (Avg. Total Working Time-Avg. Total Travel Time." "Avg. Cpu" is the average computational time to obtain a good solution, and "Std. Dev." is the standard deviation of caregivers' travel time.

Table 13. The brief results of of the COVID-19 instances

| | IZMIR | | | | | ISTANBUL | | | |
|---|---|---|---|---|---|---|---|---|---|
| Instance Clusters | Avg. Total Travel Time (min) | Std. Dev. | Avg. Total Working Time | Avg. Cpu (sec) | Instance Clusters | Avg. Travel Time (min) | Std. Dev. | Total Working Time | Avg. Cpu (sec) |
| 30SB | 44.46 | 1.84 | 437.46 | 96.12 | 40SF | 37.30 | 2.04 | 564.49 | 172.16 |
| 30SO | 53.18 | 3.19 | 449.93 | 70.76 | 40SZ | 50.54 | 3.40 | 557.22 | 147.44 |
| 30IZ | 67.04 | 5.39 | 459.45 | 77.00 | 40IS | 50.59 | 2.32 | 606.25 | 162.64 |
| 50SB | 60.16 | 1.88 | 712.92 | 179.32 | 60SF | 49.86 | 2.45 | 836.84 | 287.88 |
| 50SO | 72.22 | 1.83 | 720.15 | 162.32 | 60SZ | 66.96 | 1.55 | 831.39 | 270.64 |
| 50IZ | 88.11 | 5.49 | 739.42 | 157.20 | 60IS | 66.00 | 1.49 | 863.92 | 287.84 |
| 100IZ | 142.40 | 4.51 | 1278.76 | 695.56 | 100IS | 98.67 | 3.85 | 1420.66 | 760.56 |
| 68IZB | 87.96 | 1.56 | 1408.96 | 299.80 | 101ISF | 72.70 | 0.71 | 1403.25 | 813.40 |
| 63IZO | 74.51 | 3.40 | 1396.51 | 317.40 | 83ISZ | 91.95 | 1.37 | 1193.86 | 559.80 |
| 131IZBO | 157.31 | 4.07 | 1877.43 | 825.80 | 184ISFZ | 152.10 | 5.74 | 2590.32 | 1238.40 |
| | | | | ANKARA | | | | | |
| Instance Clusters | Avg. Travel Time (min) | Std. Dev. | Total Working Time | Avg. Cpu (sec) | Instance Clusters | Avg. Travel Time (min) | Std. Dev. | Total Working Time | Avg. Cpu (sec) |
| 75SA | 121.07 | 2.09 | 1096.61 | 423.12 | 100AN | 262.91 | 12.26 | 1604.52 | 652.32 |
| 75SC | 207.08 | 14.16 | 1177.70 | 385.20 | 149ANA | 203.23 | 2.38 | 2167.06 | 993.00 |
| 75AN | 206.17 | 8.57 | 1187.38 | 380.04 | 195ANC | 326.99 | 10.03 | 2886.16 | 1158.00 |
| 100SA | 273.94 | 9.51 | 1613.66 | 631.56 | 344ANAC | 599.56 | 17.11 | 5104.66 | 1678.80 |
| 100SC | 159.07 | 3.69 | 1489.73 | 732.28 | | | | | |

Consider the instances of Izmir in which it can be seen that the travel time increases when patients from both the neighboring districts are combinedly selected rather than dealing with them individually. This phenomenon is not seen among the instances of Istanbul and Ankara. The reason for this is that in Istanbul the land areas of both the districts are very small as compare to the land areas of other cities and the patients are located very close to each other due to high urbanization. In the case of Ankara, the land area of Çankaya individually is greater than the sum of the areas of Bornova and Bayrakli, and the patients are spread all over the districts of Ankara. This can also be observed by comparing the average travel time spent by the caregivers of 131IZ, 184IS and 344AN that are 157.31, 152.10, and 599.96 minutes, respectively.

It can also be observed that the caregivers spent the bigger portion of their total working time in giving service to their patients and not more than 18% of the total working time was spent on traveling in any of the instances. The duration of the service time of caregivers is completely dependent on the number of patients present in the respective instance. As the number of patients increases, the total service time also increases. Figure 10 demonstrates the vehicle routes of one of the 40SZ and 30SO instances, respectively. The sequence of visiting the patients is exactly delineated by the HHCSS.

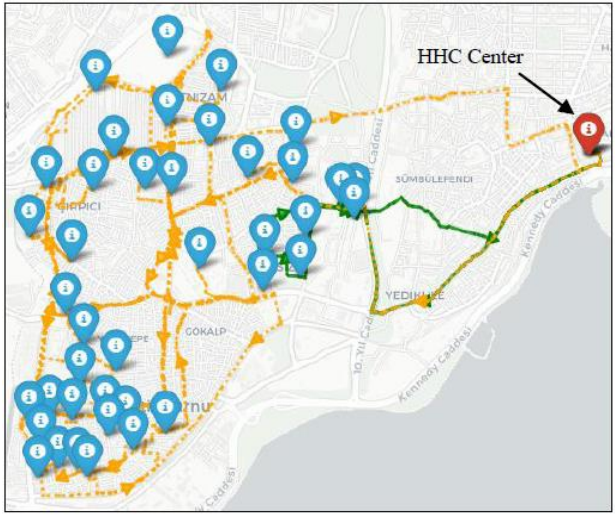 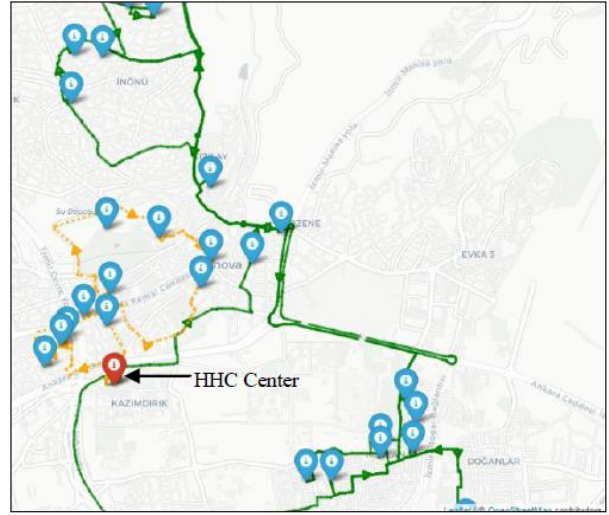

(a) Zeytinburnu neighborhood (Istanbul) with 2 caregivers covering 40 patients

(b) Bornova neighborhood (Izmir) with 2 caregivers covering 30 patients

Figure 10. The representation of the best-found road routes of the vehicles (caregivers) obtained by HHCSS. (Please see online version for the colored figures).

### 6.3    Benefits of HHCSS:

The HHCSS is designed with the current COVID-19 in mind, in which even a slight delay in controlling the pandemic can be devastating and can help in its spread. So its development aims to give priority to every COVID-19 patient without wasting much time in scheduling and routing of caregivers. It is to be expected that the use of HHCSS will save the management of the Home Healthcare center from the tedious task of scheduling and routing of caregivers and the management will also be able to shift its capabilities to other important administrative matters. These administrative matters may include satisfactory communication with patients and their families and resolving their issues regarding reservations, treatment, etc. Not only for the management unit, but this HHCS can also be very helpful for caregivers and vehicles to easily track their patients and routes. In short, it can be said that the implementation of HHCSS in HHC is to improve the quality of care rather than saving money. It is also believed that the developed HCSS through custom modifications can easily be implemented in all types of HHC clinics dealing with a variety of patients.

### 7    CONCLUSION

In this study, home healthcare scheduling and routing problem with lunch break requirement (HHSRP-LB) was considered. We specifically studied on Liu et al. (2017)'s problem because it is parallel to many real-life practices and difficult to solve. Liu et al. (2017) modeled HHSRP-LB using three-index mathematical formulation. In the model, they considered the most commonly seen features in HHSRP literature: single HHC, single period planning, time windows for patients, working period for caregivers, skill requirement and patients' preferences, and lunch break for caregivers. Therefore, the objective is designed to minimize total travel distance and the penalty cost of unvisited patients.

Because of the NP-Hardness of the problem, Liu et al. (2017) couldn't obtain optimal solutions for many problem instances using CPLEX. A branch-and-price algorithm is developed to obtain optimal solutions. However, the authors couldn't either solve some large instances within a three-hour period or obtain their optimal solutions. In practice, managers are usually willing to obtain good solutions in a very short amount of time rather than optimal solutions, especially for daily planning activities. Therefore, we developed an Adaptive Large Variable Neighborhood Search (ALNS) algorithm to provide high-quality solutions in a short amount of time for Liu et al. (2017)'s problem instances.

We developed four variants of the proposed ALNS algorithm using different local search heuristics: algorithm A0 consists of break local search and Or-opt, A1 includes only break, A2 includes only Or-Opt, and A3 does not consist of any local search heuristics. After conducting a design of the experiment, we determine the best parameter values for each variant. The ANOVA results showed that the proposed algorithms are not significantly different from each other in terms of their best solutions even though the algorithm A0 slightly presents lower cost than others on average especially for large size problem instances. According to their average run time, although the algorithm A0 requires slightly more time than algorithm A3 due to additional local search, there is no statistically difference among them for 100-patient instances. Thus, these results show that the adapted removal and insertion heuristics with lunch break position and Or-opt local searches work effectively throughout the search process without requiring additional time.

In comparison with Liu et al. (2017)'s best results, we presented that the proposed algorithms obtained about 85% of the optimal solutions. Additionally, we improved the best-known solutions for about 36 problem instances. However, we failed to obtain some of the optimal solutions but an average gap of approximately 0.03%. Therefore, especially the algorithm A0 could achieve to obtain solutions up to 35% less average route cost than Liu et al. (2017) (see Table R1_100). Moreover, the algorithm also achieved to visit up to 10 more patients than Liu et al. (2017)'s algorithm in some problem instances such as R1_100, RC1_100 and R1_50. The proposed algorithm is also shown to be a robust algorithm because the best-found solutions within five replications have a very low coefficient of variance (0.3 on average). Last, the algorithm is not computationally burdened because it needs only 54 seconds on average to solve 100-patient instances.

In addition, the Home Healthcare Decision Support System (HHCSS) has been developed to address the patients of the COVID-19 pandemic. For this, the approximate locations of COVID-19 patients present in two neighboring districts of each of the three major cities of Turkey (Ankara, Istanbul, and Izmir) is extracted from the heat map available in the mobile application "Life Fits into Home" which is developed by the Turkish Ministry of Health. The proposed ALNS algorithm A0 was implemented to find the optimal order of visiting patients considering real road distances and visualization modules.

As a result, we can conclude that the proposed algorithms, the improved solutions and the developed HHCSS in this study present significant contributions to home health care scheduling and routing literature. Moreover, the proposed algorithms can be easily modified or supported by additional local search heuristics for different types of HHSRP. For instance, the researchers may add constraint handling methods or develop specific local search heuristics for the HHSRP with synchronization or precedence constraints. Similarly, the problem could also be easily extended to consider multiple HHC or multiple period planning. Hence, the proposed ALNS algorithm could be taken as a base for further developments in future studies.

ACKNOWLEDGEMENT

This work was supported by the Scientific and Technological Research Council of Turkey (TÜBİTAK), Grant No: 217M555.

# Appendix

The notations used in Tables A1, A2 and A3 were partly explained in the manuscript in section 4.2.3. Additionally, in the column *Liu_Best*, NA indicates the unsolved problem instance by Liu et al. (2017) in these tables. Therefore, the best objective value is not known for these problems. In the colum *Alg_Best*, the italic and underlined bold values specify the worse and the improved solutions over *Liu_Best*, respectively. The other values are the same to *Liu_Best*.

Table A1. The best-found solutions by the proposed algorithms and Liu et al. (2017) for the problem instances with 30 patients.

| Instances | Liu_Best | Algorithm A0 | | | | Algorithm A1 | | | Algorithm A2 | | | Algorithm A3 | | |
|---|---|---|---|---|---|---|---|---|---|---|---|---|---|---|
| | | Alg_Best | Avg. Best Unvisited | BestAvg | AvgCpu (s) | Alg_Best | BestAvg | AvgCpu (s) | Alg_Best | BestAvg | AvgCpu (s) | Alg_Best | BestAvg | AvgCpu (s) |
| C101_30 | 1383 | 1383 | 1 | 1383.8 | 9 | 1383 | 1 | 12 | 1383 | 1 | 9 | 1383 | 1 | 7 |
| C102_30 | 344 | 344 | 0 | 348.8 | 9 | 344 | 0 | 11 | 344 | 0 | 10 | 344 | 0 | 7 |
| C103_30 | 303 | 303 | 0 | 303 | 12 | 303 | 0 | 12 | 303 | 0 | 12 | 303 | 0 | 7 |
| C104_30 | 271 | 271 | 0 | 274.2 | 14 | 271 | 0 | 14 | 271 | 0 | 14 | 271 | 0 | 11 |
| C105_30 | 378 | 378 | 0 | 378 | 7 | 378 | 0 | 9 | 378 | 0 | 8 | 378 | 0 | 7 |
| C106_30 | 1361 | 1361 | 1 | 1361 | 9 | 1361 | 1 | 12 | 1361 | 1 | 9 | 1361 | 1 | 7 |
| C107_30 | 344 | 344 | 0 | 344 | 9 | 344 | 0 | 12 | 344 | 0 | 9 | 344 | 0 | 7 |
| C108_30 | 307 | 307 | 0 | 307 | 9 | 307 | 0 | 12 | 307 | 0 | 10 | 307 | 0 | 7 |
| C109_30 | 293 | 293 | 0 | 293 | 11 | 293 | 0 | 12 | 293 | 0 | 11 | 293 | 0 | 7 |
| C201_30 | 370 | 370 | 0 | 370 | 11 | 370 | 0 | 12 | 370 | 0 | 11 | 370 | 0 | 7 |
| C202_30 | 352 | 352 | 0 | 354.4 | 14 | 352 | 0 | 18 | 352 | 0 | 14 | 352 | 0 | 12 |
| C203_30 | 349 | 349 | 0 | 353.2 | 18 | 349 | 0 | 20 | 349 | 0 | 17 | 349 | 0 | 13 |
| C204_30 | 339 | 339 | 0 | 339 | 21 | 339 | 0 | 24 | 339 | 0 | 22 | 339 | 0 | 16 |
| C205_30 | 358 | 358 | 0 | 358 | 12 | 358 | 0 | 12 | 358 | 0 | 12 | 358 | 0 | 8 |
| C206_30 | 353 | 353 | 0 | 353 | 12 | 353 | 0 | 12 | 353 | 0 | 12 | 353 | 0 | 9 |
| C207_30 | 354 | 354 | 0 | 354 | 12 | 354 | 0 | 14 | 354 | 0 | 12 | 354 | 0 | 9 |
| C208_30 | 353 | 353 | 0 | 353 | 12 | 353 | 0 | 13 | 353 | 0 | 12 | 353 | 0 | 9 |
| R101_30 | 11379 | 11379 | 11 | 11379 | 10 | 11379 | 11 | 12 | 11379 | 11 | 11 | 11379 | 11 | 8 |
| R102_30 | 6408 | 6408 | 6 | 6408 | 12 | 6408 | 6 | 12 | 6408 | 6 | 12 | 6408 | 6 | 9 |
| R103_30 | 3438 | 3438 | 3 | 3438 | 12 | 3438 | 3 | 12 | 3438 | 3 | 12 | 3438 | 3 | 8 |
| R104_30 | 1479 | 1479 | 1 | 1492.8 | 11 | 1479 | 1 | 12 | 1479 | 1 | 11 | 1479 | 1 | 7 |
| R105_30 | 7448 | 7448 | 7 | 8020 | 12 | 7448 | 7 | 12 | 7448 | 7 | 12 | 7448 | 7 | 9 |
| R106_30 | 4443 | 4443 | 4 | 4443 | 12 | 4443 | 4 | 12 | 4443 | 4 | 12 | 4443 | 4 | 9 |
| R107_30 | 1488 | 1488 | 1 | 1489.6 | 12 | 1488 | 1 | 12 | 1488 | 1 | 12 | 1488 | 1 | 7 |
| R108_30 | 508 | 508 | 0 | 508 | 11 | 508 | 0 | 12 | 508 | 0 | 11 | 508 | 0 | 7 |

Table A1. The best-found solutions by the proposed algorithms and Liu et al. (2017) for the problem instances with 30 patients (Continued).

| Instances | Liu_Best | Algorithm A0 | | | | Algorithm A1 | | | Algorithm A2 | | | Algorithm A3 | | |
|---|---|---|---|---|---|---|---|---|---|---|---|---|---|---|
| | | Alg_Best | Avg. Best Unvisited | BestAvg | AvgCpu (s) | Alg_Best | BestAvg | AvgCpu (s) | Alg_Best | BestAvg | AvgCpu (s) | Alg_Best | BestAvg | AvgCpu (s) |
| R109_30 | 4462 | 4462 | 4 | 4462 | 12 | 4462 | 4 | 12 | 4462 | 4 | 12 | 4462 | 4 | 9 |
| R110_30 | 2507 | 2507 | 2 | 2509.8 | 10 | 2507 | 2 | 10 | 2508 | 2 | 10 | *2508* | 2 | 7 |
| R111_30 | 1502 | 1502 | 1 | 1508 | 11 | 1502 | 1 | 9 | 1502 | 1 | 12 | 1502 | 1 | 7 |
| R112_30 | 501 | 501 | 0 | 501 | 9 | 501 | 0 | 9 | 501 | 0 | 9 | 501 | 0 | 7 |
| R201_30 | 522 | 522 | 0 | 522 | 11 | 522 | 0 | 11 | 522 | 0 | 11 | 522 | 0 | 7 |
| R202_30 | 472 | 472 | 0 | 472 | 12 | 472 | 0 | 12 | 472 | 0 | 12 | 472 | 0 | 7 |
| R203_30 | 457 | 457 | 0 | 457 | 16 | 457 | 0 | 14 | 457 | 0 | 16 | 457 | 0 | 12 |
| R204_30 | 431 | 431 | 0 | 431 | 18 | 431 | 0 | 17 | 431 | 0 | 17 | 431 | 0 | 14 |
| R205_30 | 479 | 479 | 0 | 479 | 12 | 479 | 0 | 12 | 479 | 0 | 12 | 479 | 0 | 7 |
| R206_30 | 467 | 467 | 0 | 467 | 12 | 467 | 0 | 12 | 467 | 0 | 12 | 467 | 0 | 11 |
| R207_30 | 444 | 444 | 0 | 444 | 16 | 444 | 0 | 16 | 444 | 0 | 16 | 444 | 0 | 12 |
| R208_30 | 416 | 416 | 0 | 416.8 | 21 | 416 | 0 | 21 | 416 | 0 | 21 | 416 | 0 | 16 |
| R209_30 | 475 | 475 | 0 | 475 | 12 | 475 | 0 | 12 | 475 | 0 | 12 | 475 | 0 | 11 |
| R210_30 | 471 | 471 | 0 | 471 | 14 | 471 | 0 | 14 | 471 | 0 | 14 | 471 | 0 | 11 |
| R211_30 | 440 | 440 | 0 | 440 | 16 | 440 | 0 | 17 | 440 | 0 | 17 | 440 | 0 | 12 |
| RC101_30 | 7467 | 7467 | 7 | 7467 | 10 | 7467 | 7 | 12 | 7467 | 7 | 10 | 7467 | 7 | 7 |
| RC102_30 | 5544 | 5544 | 5 | 5544 | 12 | 5544 | 5 | 12 | 5544 | 5 | 12 | 5544 | 5 | 7 |
| RC103_30 | 3568 | 3568 | 3 | 3778.6 | 12 | 3568 | 3 | 12 | 3568 | 3 | 12 | 3568 | 3 | 8 |
| RC104_30 | 2583 | 2583 | 2 | 2584.8 | 11 | 2583 | 2 | 11 | 2583 | 2 | 12 | 2583 | 2 | 7 |
| RC105_30 | 6515 | 6515 | 6 | 6515 | 12 | 6515 | 6 | 13 | 6515 | 6 | 12 | 6515 | 6 | 7 |
| RC106_30 | 5549 | 5549 | 5 | 5549 | 10 | 5549 | 5 | 12 | 5549 | 5 | 10 | 5549 | 5 | 7 |
| RC107_30 | 5485 | 5485 | 5 | 5485 | 12 | 5485 | 5 | 12 | 5485 | 5 | 12 | 5485 | 5 | 7 |
| RC108_30 | 3527 | *3564* | 3 | 3570.4 | 12 | *3572* | 3 | 12 | *3542* | 3 | 12 | *3542* | 3 | 7 |
| RC201_30 | 775 | 775 | 0 | 776.6 | 10 | *777* | 0 | 12 | 777 | 0 | 9 | *777* | 0 | 7 |
| RC202_30 | 648 | 648 | 0 | 648 | 12 | 648 | 0 | 12 | 648 | 0 | 12 | 648 | 0 | 7 |
| RC203_30 | 602 | 602 | 0 | 602 | 14 | 602 | 0 | 16 | 602 | 0 | 14 | 602 | 0 | 11 |
| RC204_30 | NA | **549** | 0 | 549 | 16 | **549** | 0 | 19 | **549** | 0 | 16 | **549** | 0 | 12 |
| RC205_30 | 702 | 702 | 0 | 703.6 | 12 | 702 | 0 | 12 | 702 | 0 | 12 | 702 | 0 | 7 |
| RC206_30 | 679 | 679 | 0 | 681.4 | 11 | 679 | 0 | 12 | 679 | 0 | 11 | 679 | 0 | 7 |
| RC207_30 | 615 | 615 | 0 | 618 | 15 | 615 | 0 | 18 | 615 | 0 | 12 | 615 | 0 | 11 |
| RC208_30 | 552 | 556 | 0 | 556.8 | 19 | *557* | 0 | 21 | 552 | 0 | 19 | 552 | 0 | 12 |

Table A2. The best-found solutions by the proposed algorithms and Liu et al. (2017) for the problem instances with 50 patients.

| Instances | Liu_Best | Algorithm A0 ||||| Algorithm A1 |||| Algorithm A2 |||| Algorithm A3 |||
|---|---|---|---|---|---|---|---|---|---|---|---|---|---|---|---|---|
| | | Alg_Best | Avg. Best Unvisited | BestAvg | AvgCpu (s) | Alg_Best | BestAvg | AvgCpu (s) | Alg_Best | BestAvg | AvgCpu (s) | Alg_Best | BestAvg | AvgCpu (s) |
| C101_50 | 7778 | 7778 | 7 | 7778 | 22 | 7778 | 7 | 25 | 7778 | 7 | 22 | 7778 | 7 | 17 |
| C102_50 | 1904 | 1904 | 1 | 1910.2 | 19 | 1904 | 1 | 19 | 1904 | 1 | 19 | 1904 | 1 | 12 |
| C103_50 | 768 | 768 | 0 | 779 | 19 | 768 | 0 | 19 | 768 | 0 | 19 | 768 | 0 | 14 |
| C104_50 | 637 | **_621_** | 0 | 621 | 24 | **_621_** | 0 | 29 | **_621_** | 0 | 24 | **_621_** | 0 | 19 |
| C105_50 | 4000 | 4000 | 3 | 4503 | 20 | 4000 | 3 | 24 | 4000 | 3 | 21 | 4000 | 3 | 17 |
| C106_50 | 5834 | 5834 | 5 | 5863.8 | 21 | 5834 | 5 | 24 | 5834 | 5 | 21 | 5834 | 5 | 17 |
| C107_50 | 1784 | 1784 | 1 | 1784 | 19 | 1784 | 1 | 23 | 1784 | 1 | 19 | 1784 | 1 | 14 |
| C108_50 | 824 | 824 | 0 | 830.6 | 19 | 824 | 0 | 22 | 824 | 0 | 19 | 824 | 0 | 14 |
| C109_50 | 673 | 673 | 0 | 677.4 | 19 | 673 | 0 | 24 | 673 | 0 | 19 | 673 | 0 | 16 |
| C201_50 | 813 | 813 | 0 | 815.2 | 17 | 813 | 0 | 21 | 813 | 0 | 17 | 813 | 0 | 13 |
| C202_50 | 739 | 739 | 0 | 739 | 30 | 739 | 0 | 24 | 739 | 0 | 21 | 739 | 0 | 17 |
| C203_50 | 724 | 724 | 0 | 724.8 | 31 | _725_ | 0 | 31 | _725_ | 0 | 27 | 724 | 0 | 21 |
| C204_50 | 713 | **_683_** | 0 | 691.8 | 35 | **_683_** | 0 | 35 | **_683_** | 0 | 30 | **_683_** | 0 | 24 |
| C205_50 | 800 | 800 | 0 | 802.8 | 22 | 800 | 0 | 21 | 800 | 0 | 19 | 800 | 0 | 15 |
| C206_50 | 784 | 784 | 0 | 793.4 | 21 | 784 | 0 | 23 | 784 | 0 | 19 | 784 | 0 | 16 |
| C207_50 | 779 | 779 | 0 | 779.8 | 24 | 779 | 0 | 24 | 779 | 0 | 21 | 779 | 0 | 16 |
| C208_50 | 781 | _785_ | 0 | 785 | 24 | _785_ | 0 | 24 | _785_ | 0 | 19 | _785_ | 0 | 16 |
| R101_50 | 22648 | 22648 | 22 | 22649 | 26 | 22649 | 22 | 27 | 22649 | 22 | 22 | 22649 | 22 | 17 |
| R102_50 | 14717 | 14717 | 14 | 14717 | 24 | 14717 | 14 | 24 | 14717 | 14 | 22 | 14717 | 14 | 17 |
| R103_50 | 9794 | _9821_ | 9 | _9821_ | 22 | _9821_ | 9 | 22 | _9805_ | 9 | 22 | _9805_ | 9 | 17 |
| R104_50 | 4743 | **_4739_** | 4 | 4750.8 | 22 | **_4739_** | 4 | 21 | **_4739_** | 4 | 22 | **_4739_** | 4 | 17 |
| R105_50 | 16713 | 16713 | 16 | 16713 | 22 | 16713 | 16 | 22 | 16713 | 16 | 22 | 16713 | 16 | 17 |
| R106_50 | 10797 | 10797 | 10 | 10990.6 | 22 | 10797 | 10 | 22 | 10797 | 10 | 22 | 10797 | 10 | 17 |
| R107_50 | 8757 | **_8749_** | 8 | 8752.8 | 25 | **_8749_** | 8 | 22 | **_8749_** | 8 | 22 | **_8748_** | 8 | 17 |
| R108_50 | 3773 | **_3769_** | 3 | 3771 | 22 | **_3769_** | 3 | 21 | **_3763_** | 3 | 21 | **_3763_** | 3 | 17 |
| R109_50 | 11780 | 11780 | 11 | 11785 | 22 | 11780 | 11 | 22 | 11780 | 11 | 22 | 11780 | 11 | 17 |
| R110_50 | 8808 | _8829_ | 8 | 9201.6 | 24 | _8829_ | 8 | 22 | _8821_ | 8 | 22 | _8821_ | 8 | 17 |
| R111_50 | 7810 | 7810 | 7 | 7812.2 | 23 | 7810 | 7 | 22 | _7812_ | 7 | 22 | _7812_ | 7 | 17 |
| R112_50 | 5782 | **_5765_** | 5 | 6158.8 | 24 | **_5765_** | 5 | 22 | **_5773_** | 5 | 22 | **_5773_** | 5 | 17 |

Table A2. The best-found solutions by the proposed algorithms and Liu et al. (2017) for the problem instances with 50 patients (Continued).

| Instances | Liu_Best | Algorithm A0 ||||| Algorithm A1 |||| Algorithm A2 |||| Algorithm A3 |||
|---|---|---|---|---|---|---|---|---|---|---|---|---|---|---|---|
| | | Alg_Best | Avg. Best Unvisited | BestAvg | AvgCpu (s) | Alg_Best | BestAvg | AvgCpu (s) | Alg_Best | BestAvg | AvgCpu (s) | Alg_Best | BestAvg | AvgCpu (s) |
| R201_50 | 1090 | 1090 | 0 | 1093 | 19 | 1090 | 0 | 19 | 1090 | 0 | 19 | 1090 | 0 | 14 |
| R202_50 | 999 | 999 | 0 | 1000.8 | 24 | 999 | 0 | 23 | 999 | 0 | 24 | 999 | 0 | 17 |
| R203_50 | 876 | 876 | 0 | 876 | 26 | 876 | 0 | 26 | 876 | 0 | 26 | 876 | 0 | 19 |
| R204_50 | NA | **791** | 0 | 793.6 | 28 | **791** | 0 | 28 | **791** | 0 | 28 | **791** | 0 | 22 |
| R205_50 | 984 | 984 | 0 | 984.4 | 21 | 984 | 0 | 21 | 984 | 0 | 21 | 984 | 0 | 16 |
| R206_50 | 893 | 893 | 0 | 896.2 | 24 | 893 | 0 | 24 | 893 | 0 | 24 | 893 | 0 | 19 |
| R207_50 | 833 | **828** | 0 | 828.6 | 29 | **828** | 0 | 28 | **828** | 0 | 28 | **828** | 0 | 21 |
| R208_50 | NA | **773** | 0 | 773 | 32 | **773** | 0 | 31 | **773** | 0 | 32 | **773** | 0 | 24 |
| R209_50 | 896 | 896 | 0 | 897.4 | 24 | 896 | 0 | 24 | 896 | 0 | 24 | 896 | 0 | 19 |
| R210_50 | 900 | *904* | 0 | 904 | 24 | *904* | 0 | 23 | 900 | 0 | 24 | 900 | 0 | 18 |
| R211_50 | 807 | *809* | 0 | 813.4 | 28 | 807 | 0 | 28 | 807 | 0 | 28 | 807 | 0 | 21 |
| RC101_50 | 19686 | 19686 | 19 | 19686 | 22 | 19686 | 19 | 22 | 19686 | 19 | 22 | 19686 | 19 | 17 |
| RC102_50 | 14837 | *15748* | 15 | 15748 | 22 | 14837 | 14 | 22 | *15748* | 15 | 22 | *15748* | 15 | 17 |
| RC103_50 | 12772 | 12772 | 12 | 12772 | 23 | 12772 | 12 | 22 | 12772 | 12 | 22 | 12772 | 12 | 17 |
| RC104_50 | 8869 | 8869 | 8 | 9270.4 | 22 | 8869 | 8 | 22 | 8869 | 8 | 22 | 8869 | 8 | 17 |
| RC105_50 | 17703 | 17703 | 17 | 17705 | 22 | 17703 | 17 | 22 | *17704* | 17 | 22 | 17703 | 17 | 17 |
| RC106_50 | 15859 | 15859 | 15 | 16578.2 | 22 | 15859 | 15 | 22 | *16758* | 16 | 22 | *16758* | 15 | 17 |
| RC107_50 | 12835 | 12835 | 12 | 13057.8 | 23 | *12879* | 12 | 22 | *12876* | 12 | 22 | *12836* | 12 | 17 |
| RC108_50 | 10917 | 10917 | 10 | 11651 | 24 | 10917 | 10 | 22 | *11806* | 11 | 22 | *11806* | 10 | 17 |
| RC201_50 | 1347 | 1347 | 0 | 1347 | 19 | 1347 | 0 | 17 | 1347 | 0 | 18 | 1347 | 0 | 12 |
| RC202_50 | 1243 | 1243 | 0 | 1243 | 22 | 1243 | 0 | 19 | 1243 | 0 | 21 | 1243 | 0 | 16 |
| RC203_50 | 1122 | **1117** | 0 | 1117.6 | 24 | **1117** | 0 | 23 | **1117** | 0 | 24 | **1117** | 0 | 18 |
| RC204_50 | NA | **1004** | 0 | 1009.6 | 28 | **1004** | 0 | 27 | **1004** | 0 | 28 | **1004** | 0 | 20 |
| RC205_50 | 1254 | 1254 | 0 | 1260 | 20 | 1254 | 0 | 19 | 1254 | 0 | 19 | 1254 | 0 | 15 |
| RC206_50 | 1228 | 1228 | 0 | 1228 | 20 | 1228 | 0 | 19 | 1228 | 0 | 19 | 1228 | 0 | 16 |
| RC207_50 | 1073 | 1073 | 0 | 1073 | 24 | 1073 | 0 | 24 | 1073 | 0 | 24 | 1073 | 0 | 17 |
| RC208_50 | 998 | *1015* | 0 | 1015.8 | 24 | *1016* | 0 | 24 | *1016* | 0 | 24 | *1014* | 0 | 19 |

Table A3. The best-found solutions by the proposed algorithms and Liu et al. (2017) for the problem instances with 100 patients.

| Instances | Liu_Best | Algorithm A0 | | | | Algorithm A1 | | | Algorithm A2 | | | Algorithm A3 | | |
|---|---|---|---|---|---|---|---|---|---|---|---|---|---|---|
| | | Alg_Best | Avg. Best Unvisited | BestAvg | AvgCpu (s) | Alg_Best | BestAvg | AvgCpu (s) | Alg_Best | BestAvg | AvgCpu (s) | Alg_Best | BestAvg | AvgCpu (s) |
| C101_100 | 19273 | 19273 | 17 | 20035.4 | 55 | 19273 | 17 | 57 | 19273 | 17 | 53 | 19273 | 17 | 54 |
| C102_100 | 10396 | *10398* | 8 | 10418.2 | 51 | 10396 | 8 | 49 | *10398* | 8 | 49 | 10396 | 8 | 48 |
| C103_100 | 4281 | *4294* | 2 | 4329 | 48 | *4325* | 2 | 46 | *4324* | 2 | 46 | *4320* | 2 | 46 |
| C104_100 | 1943 | **1930** | 0 | 1948.4 | 48 | **1926** | 0 | 49 | **1936** | 0 | 50 | *1952* | 0 | 50 |
| C105_100 | 13448 | **13414** | 11 | 14004.6 | 51 | *13576* | 11 | 52 | **13414** | 11 | 52 | **13414** | 11 | 53 |
| C106_100 | 12676 | *12705* | 10 | 13261.4 | 51 | *12686* | 10 | 52 | *12732* | 10 | 51 | **12665** | 10 | 53 |
| C107_100 | 7697 | *8454* | 6 | 8886.6 | 48 | *8537* | 6 | 48 | *8470* | 6 | 48 | *9357* | 7 | 50 |
| C108_100 | 5749 | **5563** | 3 | 6136.2 | 48 | *6369* | 4 | 47 | *6421* | 4 | 48 | *6396* | 4 | 48 |
| C109_100 | 2392 | **2327** | 0 | 2461.6 | 44 | **2327** | 0 | 44 | **2335** | 0 | 45 | **2318** | 0 | 45 |
| C201_100 | 2222 | 2222 | 0 | 2230.4 | 41 | 2222 | 0 | 41 | 2222 | 0 | 41 | 2222 | 0 | 41 |
| C202_100 | 2000 | 2000 | 0 | 2004.2 | 50 | 2000 | 0 | 49 | 2000 | 0 | 49 | 2000 | 0 | 51 |
| C203_100 | 1845 | 1845 | 0 | 1851.4 | 57 | 1845 | 0 | 56 | 1845 | 0 | 57 | *1848* | 0 | 57 |
| C204_100 | 1759 | *1768* | 0 | 1779.4 | 63 | *1760* | 0 | 63 | **1754** | 0 | 64 | **1754** | 0 | 65 |
| C205_100 | 2111 | 2111 | 0 | 2111 | 41 | 2111 | 0 | 42 | 2111 | 0 | 42 | 2111 | 0 | 43 |
| C206_100 | 2087 | *2089* | 0 | 2095.8 | 43 | *2089* | 0 | 43 | *2089* | 0 | 44 | *2089* | 0 | 45 |
| C207_100 | 2057 | 2057 | 0 | 2063.2 | 45 | 2057 | 0 | 45 | *2058* | 0 | 47 | *2058* | 0 | 47 |
| C208_100 | 2015 | *2018* | 0 | 2029.2 | 45 | *2018* | 0 | 46 | *2018* | 0 | 46 | *2018* | 0 | 47 |
| R101_100 | 41225 | *41229* | 40 | 41233.8 | 56 | *41229* | 40 | 58 | 41225 | 40 | 57 | 41225 | 40 | 61 |
| R102_100 | 26496 | 26496 | 25 | 26682.8 | 55 | 26496 | 25 | 57 | 26496 | 25 | 57 | 26496 | 25 | 58 |
| R103_100 | 16582 | **16534** | 15 | 17148.4 | 53 | **16534** | 15 | 54 | **16534** | 15 | 54 | **16534** | 15 | 56 |
| R104_100 | 10575 | *10600* | 9 | 11565.4 | 51 | *10600* | 9 | 52 | **10541** | 9 | 55 | **10543** | 9 | 55 |
| R105_100 | 30412 | *30442* | 29 | 31217.2 | 58 | *30442* | 29 | 58 | *30428* | 29 | 58 | *30428* | 29 | 60 |
| R106_100 | 18686 | *19593* | 18 | 20573.6 | 56 | *20550* | 19 | 55 | *19596* | 18 | 55 | *19596* | 18 | 57 |
| R107_100 | 13560 | **12596** | 11 | 13003 | 54 | **12596** | 11 | 53 | **12574** | 11 | 54 | **12574** | 11 | 55 |
| R108_100 | 7622 | **7576** | 6 | 8570.4 | 54 | *8581* | 7 | 52 | **7576** | 6 | 52 | *8586* | 7 | 54 |
| R109_100 | 20587 | *21506* | 20 | 21725.8 | 57 | *21506* | 20 | 57 | *21516* | 20 | 57 | *21510* | 20 | 58 |
| R110_100 | 16572 | *16581* | 15 | 17374.4 | 58 | *16581* | 15 | 55 | *16524* | 15 | 57 | *16570* | 15 | 57 |
| R111_100 | 13584 | **13571** | 12 | 14201.2 | 54 | **13571** | 12 | 53 | **13541** | 12 | 53 | *14550* | 13 | 55 |

Table A3. The best-found solutions by the proposed algorithms and Liu et al. (2017) for the problem instances with 100 patients (Continued).

| Instances | Liu_Best | Algorithm A0 | | | | Algorithm A1 | | | Algorithm A2 | | | Algorithm A3 | | |
|---|---|---|---|---|---|---|---|---|---|---|---|---|---|---|
| | | Alg_Best | Avg. Best Unvisited | BestAvg | AvgCpu (s) | Alg_Best | BestAvg | AvgCpu (s) | Alg_Best | BestAvg | AvgCpu (s) | Alg_Best | BestAvg | AvgCpu (s) |
| R112_100 | 11530 | **10559** | 9 | 12758.4 | 56 | *11538* | 10 | 55 | *11534* | 10 | 56 | **10548** | 9 | 57 |
| R201_100 | 2011 | 2011 | 0 | 2012.6 | 43 | 2011 | 0 | 41 | 2011 | 0 | 42 | *2012* | 0 | 44 |
| R202_100 | 1794 | **1793** | 0 | 1794.8 | 47 | **1793** | 0 | 47 | **1793** | 0 | 48 | **1793** | 0 | 48 |
| R203_100 | 1716 | **1615** | 0 | 1619.2 | 54 | **1621** | 0 | 54 | **1615** | 0 | 54 | **1621** | 0 | 54 |
| R204_100 | NA | **1547** | 0 | 1550.2 | 62 | **1543** | 0 | 64 | **1543** | 0 | 62 | **1547** | 0 | 66 |
| R205_100 | 1890 | **1821** | 0 | 1826.8 | 48 | **1811** | 0 | 48 | **1798** | 0 | 47 | **1811** | 0 | 53 |
| R206_100 | 1765 | **1686** | 0 | 1700.4 | 52 | **1686** | 0 | 52 | **1686** | 0 | 52 | **1681** | 0 | 59 |
| R207_100 | 1718 | **1589** | 0 | 1596 | 58 | **1593** | 0 | 58 | **1596** | 0 | 59 | **1593** | 0 | 63 |
| R208_100 | NA | **1516** | 0 | 1518.8 | 66 | 1516 | 0 | 66 | 1516 | 0 | 66 | 1516 | 0 | 72 |
| R209_100 | 1770 | **1667** | 0 | 1689.8 | 53 | **1675** | 0 | 53 | **1663** | 0 | 55 | **1675** | 0 | 57 |
| R210_100 | 1719 | **1645** | 0 | 1665 | 53 | **1645** | 0 | 52 | **1661** | 0 | 54 | **1645** | 0 | 54 |
| R211_100 | 1609 | **1547** | 0 | 1562 | 58 | **1558** | 0 | 58 | **1553** | 0 | 58 | **1548** | 0 | 58 |
| RC101_100 | 37507 | 37507 | 36 | 37706.8 | 57 | 37507 | 36 | 58 | 37507 | 36 | 58 | 37507 | 36 | 61 |
| RC102_100 | 28703 | 28703 | 27 | 29114.6 | 58 | *28704* | 27 | 57 | 28703 | 27 | 59 | 28703 | 27 | 60 |
| RC103_100 | 20778 | *21707* | 20 | 22106.8 | 56 | 21707 | 20 | 57 | *21738* | 20 | 59 | *21738* | 20 | 62 |
| RC104_100 | 16711 | *16713* | 15 | 17347 | 57 | *16713* | 15 | 57 | *16750* | 15 | 57 | *16736* | 15 | 62 |
| RC105_100 | 30823 | 30823 | 29 | 31564.4 | 65 | 30823 | 29 | 60 | *31753* | 30 | 59 | 30823 | 29 | 64 |
| RC106_100 | 28517 | *28527* | 27 | 28587.4 | 63 | 28527 | 27 | 57 | *28531* | 27 | 57 | *28551* | 27 | 64 |
| RC107_100 | 21756 | *22672* | 21 | 23706 | 63 | 22672 | 21 | 58 | *23637* | 22 | 59 | *23637* | 22 | 64 |
| RC108_100 | 19744 | *20715* | 19 | 21537.6 | 63 | **19739** | 18 | 60 | *19750* | 18 | 60 | *19750* | 18 | 65 |
| RC201_100 | 2516 | 2516 | 0 | 2519.4 | 49 | 2516 | 0 | 43 | 2516 | 0 | 43 | 2516 | 0 | 48 |
| RC202_100 | 2353 | **2299** | 0 | 2318.2 | 53 | **2299** | 0 | 48 | **2299** | 0 | 48 | **2299** | 0 | 48 |
| RC203_100 | 2135 | **2078** | 0 | 2088.2 | 60 | **2078** | 0 | 54 | **2078** | 0 | 55 | **2081** | 0 | 54 |
| RC204_100 | NA | **1923** | 0 | 1926 | 69 | **1923** | 0 | 65 | **1925** | 0 | 63 | **1917** | 0 | 64 |
| RC205_100 | 2316 | 2316 | 0 | 2316 | 54 | 2316 | 0 | 46 | 2316 | 0 | 46 | 2316 | 0 | 46 |
| RC206_100 | 2199 | 2199 | 0 | 2217.4 | 52 | *2221* | 0 | 47 | *2201* | 0 | 47 | *2201* | 0 | 47 |
| RC207_100 | 2084 | **2044** | 0 | 2053.6 | 57 | **2044** | 0 | 53 | **2056** | 0 | 53 | **2056** | 0 | 52 |
| RC208_100 | 2006 | **1929** | 0 | 1938 | 66 | **1929** | 0 | 59 | **1930** | 0 | 60 | **1930** | 0 | 58 |

Table A4. The comparison of the best solutions presented by Liu et al. (2017) and provided by algorithm A0 in details.

| Instance Classes | Tot. No. Instances | Liu et al. (2017) | | | | | | | Algorithm A0 | | | |
|---|---|---|---|---|---|---|---|---|---|---|---|---|
| | | *Unsolved* | *Tgap %* | Solved No | *OptVal Avg* | *OptVal Std* | *Avg. Unvisited* | *TCPU (s)* | *BestAvg* | *BestStd* | *Avg.Best Unvisited* | *Avg.Best CPU (s)* |
| C1_30 | 9 | 0 | 0.0 | 9 | 553.8 | 438.4 | 0.2 | 49.3 | 553.78 | 438.4 | 0.2 | 9.89 |
| C2_30 | 8 | 0 | 0.0 | 8 | 353.5 | 8.1 | 0.0 | 1517.7 | 353.50 | 8.1 | 0 | 14.00 |
| R1_30 | 12 | 0 | 0.0 | 12 | 3796.9 | 3135.0 | 3.3 | 226.0 | 3796.92 | 3135.0 | 3.3 | 11.17 |
| R2_30 | 11 | 0 | 0.0 | 11 | 461.3 | 27.2 | 0.0 | 1178.8 | 461.27 | 27.2 | 0 | 14.55 |
| RC1_30 | 8 | 1 | 5.3 | 7 | 5244.4 | 1544.0 | 4.7 | 923.6 | 5034.38 | 1547.5 | 4.5 | 11.38 |
| RC2_30 | 8 | 1 | --- | 7 | 653.3 | 67.9 | 0.0 | 3416.8 | 640.75 | 71.6 | 0 | 13.63 |
| C1_50 | 9 | 1 | 3.5 | 8 | 2945.6 | 2491.2 | 2.1 | 546.5 | 2687.33 | 2459.7 | 1.9 | 20.22 |
| C2_50 | 8 | 1 | 7.2 | 7 | 774.3 | 29.5 | 0.0 | 1293.0 | 763.38 | 41.1 | 0 | 25.50 |
| R1_50 | 12 | 4 | 14.0 | 8 | 12883.4 | 4621.1 | 12.1 | 1741.5 | 10511.42 | 5165.2 | 9.8 | 23.17 |
| R2_50 | 11 | 4 | 0.7 | 7 | 948.3 | 73.0 | 0.0 | 1076.5 | 894.82 | 92.8 | 0 | 25.36 |
| RC1_50 | 8 | 0 | 0.0 | 8 | 14184.8 | 3328.4 | 13.4 | 565.8 | 14298.63 | 3364.1 | 13.5 | 22.50 |
| RC2_50 | 8 | 2 | 4.4 | 6 | 1190.5 | 118.0 | 0.0 | 1317.3 | 1160.13 | 117.3 | 0 | 22.63 |
| C1_100 | 9 | 5 | 7.4 | 4 | 10411.8 | 5556.0 | 8.0 | 1970.2 | 8706.44 | 5464.0 | 6.3 | 49.33 |
| C2_100 | 8 | 1 | 2.6 | 7 | 2048.1 | 107.2 | 0.0 | 1765.1 | 2013.75 | 136.7 | 0 | 48.13 |
| R1_100 | 12 | 8 | 21.9 | 4 | 29204.8 | 8122.9 | 27.8 | 2547.0 | 18940.25 | 9321.0 | 17.4 | 55.17 |
| R2_100 | 11 | 10 | 6.8 | 1 | 2011.0 | 0.0 | 0.0 | 4418.9 | 1676.09 | 140.9 | 0 | 54.00 |
| RC1_100 | 8 | 5 | 6.0 | 3 | 32344.3 | 3751.8 | 30.7 | 2769.1 | 25920.88 | 6255.0 | 24.3 | 60.25 |
| RC2_100 | 8 | 5 | 5.3 | 3 | 2343.7 | 130.9 | 0.0 | 4020.8 | 2163.00 | 194.1 | 0 | 57.50 |

In Table A4, *Unsolved* indicates the number of instances that were not solved optimally by Liu et al. (2017). *Tgap* presents the average gap between the lower and upper bounds of *Liu_Best*. *OptVal Avg*, *OptVal Std* refer the mean and the standard deviations of the *Liu_Best* in the given set of problem instances. Similarly, *Avg. Unvisited* indicate the average number of unvisited patients in *Liu_Best* for the specified instance class. *TCPU* is the average computational time of *Liu_Best*s. To make an accurate comparison, *BestAvg* and *BestStd* defines the mean and the standard deviations of the *Best* of the algorithm A0. *Avg. Best Unvisited* and *Avg. Best CPU* indicate the average number of unvisited patients and run time in *Best* provided by the algorithm A0.

The notations used in Tables A5, A6, and A7 are described as in the followings: "Best Total Travel Time" shows the best found solution that minimizes the total travel time spent by caregivers, "Avg. Total Travel Time" is the average of the best total travel times obtained in each replication, "Max Total Travel Time" is the worst solution, "Gap1" is the percentage gap between the best and the average solutions, and "Gap2" is the percentage gap between the best and the worst solutions. Last, "Avg. CPU" is the average of computational time spent for solving all corresponding instances and replications.

Table A5. The detail of COVID-19 patients results of Izmir

| Instances | Best Total Travel Time | Avg. Total Travel Time | Max Total Travel Time | GAP1 (%) | GAP2 (%) | Avg. CPU |
|---|---|---|---|---|---|---|
| 30SB1 | 45.74 | 45.82 | 45.93 | 0.18 | 0.25 | 96.4 |
| 30SB2 | 44.34 | 44.51 | 44.61 | 0.39 | 0.22 | 86.8 |
| 30SB3 | 42.94 | 42.94 | 42.94 | 0.00 | 0.00 | 97.6 |
| 30SB4 | 42.26 | 42.26 | 42.27 | 0.01 | 0.02 | 100.2 |
| 30SB5 | 47.04 | 47.24 | 47.42 | 0.41 | 0.37 | 99.6 |
| 30SO1 | 54.64 | 54.64 | 54.64 | 0.00 | 0.00 | 65 |
| 30SO2 | 58.24 | 58.25 | 58.28 | 0.02 | 0.05 | 66.6 |
| 30SO3 | 49.25 | 49.28 | 49.33 | 0.07 | 0.10 | 56.6 |
| 30SO4 | 50.95 | 50.95 | 50.95 | 0.00 | 0.00 | 87.8 |
| 30SO5 | 52.82 | 52.83 | 52.86 | 0.01 | 0.06 | 77.8 |
| 30IZ1 | 68.20 | 68.41 | 68.98 | 0.30 | 0.80 | 74.6 |
| 30IZ2 | 62.60 | 62.84 | 63.17 | 0.39 | 0.51 | 86 |
| 30IZ3 | 70.03 | 70.50 | 71.11 | 0.68 | 0.85 | 67.4 |
| 30IZ4 | 74.57 | 74.67 | 74.82 | 0.14 | 0.20 | 69.2 |
| 30IZ5 | 59.83 | 59.86 | 59.87 | 0.05 | 0.02 | 87.8 |
| 50SB1 | 60.97 | 61.32 | 61.75 | 0.56 | 0.70 | 183.8 |
| 50SB2 | 61.47 | 61.68 | 61.93 | 0.34 | 0.40 | 178.2 |
| 50SB3 | 60.59 | 60.80 | 61.14 | 0.34 | 0.56 | 182.8 |
| 50SB4 | 56.82 | 57.30 | 57.80 | 0.84 | 0.85 | 184.8 |
| 50SB5 | 60.96 | 61.20 | 61.52 | 0.37 | 0.52 | 167 |
| 50SO1 | 72.07 | 72.52 | 73.03 | 0.62 | 0.68 | 162.8 |
| 50SO2 | 73.90 | 74.20 | 74.73 | 0.40 | 0.71 | 164.6 |
| 50SO3 | 71.99 | 72.23 | 72.43 | 0.32 | 0.27 | 160.4 |
| 50SO4 | 69.87 | 70.46 | 70.98 | 0.84 | 0.72 | 167.6 |
| 50SO5 | 73.29 | 73.69 | 74.36 | 0.54 | 0.88 | 156.2 |
| 50IZ1 | 87.10 | 87.87 | 88.75 | 0.88 | 0.99 | 158.6 |
| 50IZ2 | 93.47 | 94.41 | 95.59 | 0.98 | 1.20 | 155.2 |
| 50IZ3 | 87.58 | 88.19 | 88.99 | 0.70 | 0.89 | 148.8 |
| 50IZ4 | 92.83 | 93.36 | 94.32 | 0.56 | 1.01 | 159.4 |
| 50IZ5 | 79.59 | 79.87 | 80.16 | 0.35 | 0.36 | 164 |
| 100IZ1 | 144.00 | 146.84 | 150.44 | 1.93 | 2.39 | 668.8 |
| 100IZ2 | 142.99 | 145.70 | 149.09 | 1.86 | 2.27 | 719.4 |
| 100IZ3 | 141.33 | 144.62 | 147.60 | 2.27 | 2.01 | 709.8 |
| 100IZ4 | 144.40 | 147.19 | 149.54 | 1.89 | 1.57 | 704.6 |
| 100IZ5 | 139.29 | 141.54 | 143.52 | 1.59 | 1.37 | 675.2 |
| 68IZB | 87.96 | 88.80 | 89.72 | 0.95 | 1.01 | 299.8 |
| 63IZO | 74.51 | 75.17 | 76.45 | 0.89 | 1.68 | 317.4 |
| 131IZBO | 157.31 | 161.35 | 165.12 | 2.51 | 2.27 | 825.8 |

Table A6. The detail of COVID-19 patients results of Istanbul

| Instances | Best Total Travel Time | Avg. Total Travel Time | Max Total Travel Time | GAP1 (%) | GAP2 (%) | Avg. CPU |
|---|---|---|---|---|---|---|
| 40SF1 | 35.90 | 35.92 | 35.99 | 0.07 | 0.18 | 176.4 |
| 40SF2 | 39.80 | 39.95 | 40.13 | 0.37 | 0.44 | 166.0 |
| 40SF3 | 39.53 | 39.57 | 39.72 | 0.09 | 0.36 | 163.2 |
| 40SF4 | 35.35 | 35.36 | 35.41 | 0.04 | 0.15 | 191.4 |
| 40SF5 | 35.95 | 36.07 | 36.28 | 0.33 | 0.56 | 163.8 |
| 40SZ1 | 52.66 | 52.74 | 52.77 | 0.17 | 0.05 | 147.0 |
| 40SZ2 | 51.20 | 51.25 | 51.42 | 0.09 | 0.34 | 143.6 |
| 40SZ3 | 51.22 | 51.25 | 51.28 | 0.06 | 0.05 | 148.6 |
| 40SZ4 | 44.16 | 44.26 | 44.42 | 0.23 | 0.36 | 150.4 |
| 40SZ5 | 53.48 | 53.64 | 53.88 | 0.30 | 0.45 | 147.6 |
| 40IS1 | 48.52 | 48.52 | 48.52 | 0.00 | 0.00 | 151.0 |
| 40IS2 | 48.26 | 48.38 | 48.52 | 0.25 | 0.30 | 176.6 |
| 40IS3 | 52.39 | 52.50 | 52.80 | 0.22 | 0.58 | 162.8 |
| 40IS4 | 49.98 | 50.12 | 50.66 | 0.26 | 1.01 | 171.0 |
| 40IS5 | 53.79 | 53.79 | 53.82 | 0.01 | 0.05 | 151.8 |
| 60SF1 | 47.08 | 47.50 | 47.83 | 0.87 | 0.69 | 293.6 |
| 60SF2 | 52.51 | 52.95 | 53.23 | 0.84 | 0.53 | 286.4 |
| 60SF3 | 52.26 | 52.53 | 53.09 | 0.52 | 1.01 | 288.0 |
| 60SF4 | 47.77 | 47.87 | 48.11 | 0.21 | 0.50 | 285.2 |
| 60SF5 | 49.68 | 50.04 | 50.35 | 0.72 | 0.61 | 286.2 |
| 60SZ1 | 68.36 | 69.00 | 69.68 | 0.93 | 0.96 | 276.0 |
| 60SZ2 | 66.17 | 66.37 | 66.70 | 0.31 | 0.49 | 266.6 |
| 60SZ3 | 65.70 | 65.98 | 66.41 | 0.42 | 0.64 | 276.6 |
| 60SZ4 | 67.04 | 67.38 | 67.82 | 0.50 | 0.66 | 257.6 |
| 60SZ5 | 67.54 | 67.92 | 68.30 | 0.54 | 0.56 | 276.4 |
| 60IS1 | 67.43 | 68.06 | 69.08 | 0.91 | 1.46 | 287.6 |
| 60IS2 | 65.42 | 65.95 | 66.49 | 0.81 | 0.81 | 280.4 |
| 60IS3 | 65.90 | 66.28 | 66.70 | 0.56 | 0.63 | 285.4 |
| 60IS4 | 65.31 | 65.65 | 66.08 | 0.53 | 0.65 | 278.2 |
| 60IS5 | 65.92 | 66.17 | 66.42 | 0.38 | 0.36 | 307.6 |
| 100IS1 | 98.28 | 100.94 | 103.60 | 2.62 | 2.59 | 760.4 |
| 100IS2 | 101.04 | 103.30 | 105.61 | 2.18 | 2.19 | 747.2 |
| 100IS3 | 98.02 | 99.49 | 101.22 | 1.48 | 1.70 | 753.4 |
| 100IS4 | 96.28 | 98.72 | 101.23 | 2.50 | 2.47 | 776.4 |
| 100IS5 | 99.75 | 102.28 | 105.35 | 2.44 | 2.88 | 765.4 |
| 101ISF | 72.70 | 74.77 | 77.31 | 2.76 | 3.27 | 813.4 |
| 83ISZ | 91.95 | 92.97 | 94.01 | 1.09 | 1.11 | 559.8 |
| 184ISFZ | 152.10 | 156.85 | 162.26 | 3.05 | 3.33 | 1238.4 |

Table A7. The detail of COVID-19 patients results of Ankara

| Instances | Best Total Travel Time | Avg. Total Travel Time | Max Total Travel Time | GAP1 (%) | GAP2 (%) | Avg. CPU |
|---|---|---|---|---|---|---|
| 75SA1 | 120.25 | 122.01 | 124.26 | 1.44 | 1.81 | 421.4 |
| 75SA2 | 122.00 | 123.09 | 124.69 | 0.89 | 1.27 | 421.2 |
| 75SA3 | 120.85 | 123.21 | 125.62 | 1.90 | 1.91 | 426.2 |
| 75SA4 | 119.99 | 121.14 | 122.52 | 0.95 | 1.13 | 404.4 |
| 75SA5 | 122.26 | 125.00 | 127.28 | 2.19 | 1.79 | 442.4 |
| 75SC1 | 201.84 | 203.80 | 205.80 | 0.93 | 0.97 | 393.0 |
| 75SC2 | 207.52 | 209.77 | 212.33 | 1.05 | 1.18 | 383.2 |
| 75SC3 | 213.06 | 215.20 | 217.57 | 0.98 | 1.07 | 354.8 |
| 75SC4 | 206.09 | 210.34 | 217.36 | 2.03 | 3.18 | 380.0 |
| 75SC5 | 206.92 | 209.78 | 213.11 | 1.32 | 1.49 | 415.0 |
| 75AN1 | 200.78 | 203.40 | 205.53 | 1.28 | 1.03 | 391.6 |
| 75AN2 | 207.54 | 210.77 | 215.80 | 1.54 | 2.26 | 378.4 |
| 75AN3 | 205.85 | 208.15 | 211.42 | 1.11 | 1.56 | 379.0 |
| 75AN4 | 204.83 | 206.96 | 210.02 | 1.03 | 1.47 | 379.8 |
| 75AN5 | 211.84 | 213.55 | 215.21 | 0.80 | 0.77 | 371.4 |
| 100SA1 | 277.94 | 281.94 | 286.33 | 1.43 | 1.54 | 624.0 |
| 100SA2 | 271.93 | 276.74 | 281.14 | 1.73 | 1.56 | 607.2 |
| 100SA3 | 277.09 | 283.95 | 288.56 | 2.41 | 1.58 | 592.0 |
| 100SA4 | 263.88 | 267.96 | 272.11 | 1.52 | 1.51 | 616.4 |
| 100SA5 | 278.88 | 284.93 | 290.17 | 2.13 | 1.80 | 718.2 |
| 100SC1 | 161.74 | 165.75 | 171.57 | 2.42 | 3.39 | 733.8 |
| 100SC2 | 155.45 | 160.09 | 164.98 | 2.88 | 2.97 | 745.6 |
| 100SC3 | 160.53 | 163.57 | 166.40 | 1.86 | 1.69 | 732.0 |
| 100SC4 | 158.31 | 162.42 | 166.00 | 2.54 | 2.14 | 703.0 |
| 100SC5 | 159.35 | 162.52 | 164.52 | 1.95 | 1.21 | 747.0 |
| 100AN1 | 262.29 | 271.72 | 278.99 | 3.42 | 2.59 | 659.2 |
| 100AN2 | 253.55 | 259.77 | 266.14 | 2.40 | 2.40 | 657.4 |
| 100AN3 | 256.61 | 263.61 | 272.14 | 2.66 | 3.11 | 667.2 |
| 100AN4 | 263.24 | 268.65 | 272.27 | 2.04 | 1.34 | 652.8 |
| 100AN5 | 278.86 | 286.01 | 294.80 | 2.48 | 3.00 | 625.0 |
| 149ANA | 203.23 | 209.99 | 217.23 | 3.20 | 3.34 | 993.0 |
| 195ANC | 326.99 | 336.36 | 347.94 | 2.78 | 3.30 | 1158.0 |
| 344ANAC | 599.56 | 623.35 | 650.11 | 3.83 | 4.12 | 1678.8 |